\documentclass[sigconf,10pt]{acmart}

\pdfoutput=1

\usepackage{enumitem}
\usepackage{framed}
\usepackage{cprotect}
\usepackage{enumitem}
\usepackage{listings}
\usepackage{amstext}
\usepackage{amstext}
\usepackage{pdfpages}
\usepackage{alltt}
\usepackage{epstopdf}
\usepackage{xspace,colortbl}
\usepackage{multirow}
\usepackage{amssymb}
\usepackage{fmtcount}
\usepackage{amsfonts}
\usepackage{xspace}
\usepackage{amsmath}
\usepackage{multirow}
\usepackage[mathscr]{eucal}
\usepackage{colortbl}
\usepackage{lipsum}

\usepackage{amsmath,amssymb}
\usepackage[linesnumbered, ruled,vlined]{algorithm2e}

\usepackage{caption}
\usepackage{graphicx}

\usepackage{bm}
\usepackage[nospace]{cite}
\usepackage{csquotes}
\usepackage{enumitem}

\usepackage{courier}

\usepackage{balance}

\usepackage{tikz}
\usetikzlibrary{calc,positioning,shapes.geometric}

\usepackage{verbatim}
\usepackage{subcaption}
\newsavebox{\imagebox}
\usepackage{booktabs}
\newcommand{\ra}[1]{\renewcommand{\arraystretch}{#1}}
\newcommand{\secref}[1]{\S\ref{#1}}

\DeclareMathOperator*{\argmax}{arg\,max}
\newcommand{\sys}{\textsf{DQ}\xspace}

\begin{document}

\newtheorem{problem}{Problem}

\newcommand{\detectlib}{\texttt{IsoDetect}\xspace}
\newcommand{\company}{\texttt{Company X}\xspace}
\newcommand{\cond}{\textrm{pred}\xspace}
\newcommand{\dataset}{data set\xspace}
\newcommand{\datasets}{data sets\xspace}
\newcommand{\spview}{\textsf{SPView}\xspace}
\newcommand{\fjview}{\textsf{FJView}\xspace}
\newcommand{\aggview}{\textsf{AggView}\xspace}
\newcommand{\hashfunc}[1]{\textsf{hash}(#1)\xspace}
\newcommand{\hashop}{\textsf{hash}\xspace}
\newcommand{\nsc}{\textsf{NormalizedSC}\xspace}
\newcommand{\rsc}{\textsf{RawSC}\xspace}

\newcommand{\avgfunc}{\ensuremath{\texttt{avg} }\xspace}
\newcommand{\maxfunc}{\ensuremath{\texttt{max} }\xspace}
\newcommand{\minfunc}{\ensuremath{\texttt{min} }\xspace}
\newcommand{\histfunc}{\ensuremath{\texttt{histogram\_numeric} }\xspace}
\newcommand{\countfunc}{\ensuremath{\texttt{count}}\xspace}
\newcommand{\sumfunc}{\ensuremath{\texttt{sum} }\xspace}
\newcommand{\varfunc}{\ensuremath{\texttt{var} }\xspace}
\newcommand{\stdfunc}{\ensuremath{\texttt{std} }\xspace}
\newcommand{\covfunc}{\ensuremath{\texttt{cov} }\xspace}
\newcommand{\corrfunc}{\ensuremath{\texttt{corr} }\xspace}
\newcommand{\medfunc}{\ensuremath{\texttt{median} }\xspace}
\newcommand{\percfunc}{\ensuremath{\texttt{percentile} }\xspace}
\newcommand{\havingfunc}{\ensuremath{\texttt{HAVING} }\xspace}
\newcommand{\selectfunc}{\ensuremath{\texttt{select} }\xspace}
\newcommand{\ratio}{\ensuremath{\rho }\xspace}

\newcommand{\insertion}{\ensuremath{\texttt{INSERT} }\xspace}
\newcommand{\update}{\ensuremath{\texttt{UPDATE} }\xspace}
\newcommand{\delete}{\ensuremath{\texttt{DELETE} }\xspace}

\newcommand{\reminder}[1]{{{\textcolor{magenta}{\{\{\bf #1\}\}}}\xspace}}
\newcommand{\ewu}[1]{{{\textcolor{blue}{\{\{\bf ewu:\} #1\}}}\xspace}}
\newcommand{\mps}[1]{{{\textcolor{red}{\{\{\bf meelap:\} #1\}}}\xspace}}
\newcommand{\stitle}[1]{\smallskip\noindent\textbf{#1: }}
\newcommand{\ititle}[1]{\smallskip\noindent\textit{#1: }}
\newcommand{\btitle}[1]{\smallskip\noindent\textbf{#1}}

\definecolor{light-gray}{gray}{0.95}
\definecolor{mid-gray}{gray}{0.85}
\definecolor{green}{RGB}{0,176,80}
\definecolor{darkred}{rgb}{0.7,0.25,0.25}
\definecolor{darkgreen}{rgb}{0.15,0.55,0.15}
\definecolor{darkblue}{rgb}{0.1,0.1,0.5}
\definecolor{orange}{RGB}{237,125,49}
\definecolor{blue}{RGB}{68,114,196}
\definecolor{pop}{RGB}{0,21,245}

\newcommand{\white}[1]{{\textcolor{white}{#1}\xspace}}
\newcommand{\blue}[1]{{\textcolor{blue}{{\bf #1}}\xspace}}
\newcommand{\orange}[1]{{\textcolor{orange}{{\bf #1}}\xspace}}
\newcommand{\pop}[1]{{\textcolor{pop}{{\textit{\textbf{#1}}}}\xspace}}
\newcommand{\red}[1]{\textcolor{red}{#1}}
\newcommand{\green}[1]{\textcolor{green}{#1}}
\newcommand{\gray}[1]{\textcolor{light-gray}{#1}}

\newcommand{\jmh}[1]{ {\color{red}jmh: #1}}
\newcommand{\ion}[1]{ {\color{green}ion: #1}}
\newcommand{\zongheng}[1]{ {\color{cyan}zongheng: #1}}

\newcommand{\specialcell}[2][c]{%
  \begin{tabular}[#1]{@{}c@{}}#2\end{tabular}}

\def\ojoin{\setbox0=\hbox{$\bowtie$}%
  \rule[-.02ex]{.25em}{.4pt}\llap{\rule[\ht0]{.25em}{.4pt}}}
\def\leftouterjoin{\mathbin{\ojoin\mkern-5.8mu\bowtie}}
\def\rightouterjoin{\mathbin{\bowtie\mkern-5.8mu\ojoin}}
\def\fullouterjoin{\mathbin{\ojoin\mkern-5.8mu\bowtie\mkern-5.8mu\ojoin}}

\setlength{\belowcaptionskip}{-3pt}

\pagestyle{plain}

\title{Learning to Optimize Join Queries With Deep Reinforcement Learning}


\author{\normalsize Sanjay Krishnan$^{1,2}$, Zongheng Yang$^1$, Ken Goldberg$^1$, Joseph M. Hellerstein$^1$, Ion Stoica$^1$  \\
{\normalsize $^1$RISELab, UC Berkeley~~~~~~~$^2$Computer Science, University of Chicago} \\
{\normalsize skr@cs.uchicago.edu ~~~~~~ \{zongheng, goldberg, hellerstein, istoica\}@berkeley.edu}\\
}

\fontsize{10pt}{12pt} 
\selectfont

\begin{abstract}
Exhaustive enumeration of all possible join orders is often avoided, and most optimizers leverage heuristics to prune the search space. The design and implementation of heuristics are well-understood when the cost model is roughly linear, and we find that these heuristics can be significantly suboptimal when there are non-linearities in cost. Ideally, instead of a fixed heuristic, we would want a strategy to guide the search space in a more data-driven way---tailoring the search to a specific dataset and query workload. Recognizing the link between classical Dynamic Programming enumeration methods and recent results in Reinforcement Learning (RL), we propose a new method for learning optimized join search strategies.  We present our RL-based DQ optimizer, which currently optimizes select-project-join blocks. We implement three versions of DQ to illustrate the ease of integration into existing DBMSes: (1) A version built on top of Apache Calcite, (2) a version integrated into PostgreSQL, and (3) a version integrated into SparkSQL. 
Our extensive evaluation shows that DQ achieves plans with optimization costs and query execution times competitive with the native query optimizer in each system, but can execute significantly faster after learning (often by orders of magnitude). 
\end{abstract}

\maketitle


\section{Introduction}\label{intro}\sloppy
Join optimization has been studied for
more than four decades~\citep{selinger1979access}
and continues to be an active area of research~\citep{trummer2017solving,neumann2018adaptive,marcus2018deep}.  The problem's combinatorial complexity leads to the ubiquitous use of \emph{heuristics}.  For example, classical System R-style dynamic programs often restrict their search space to certain shapes (e.g., ``left-deep'' plans).  Query optimizers sometimes apply further heuristics to large join queries using genetic~\citep{postgres-genetic} or randomized~\citep{neumann2018adaptive} algorithms.  
In edge cases, these heuristics can break down (by definition), which results in poor plans~\citep{leis2015good}.  

In light of recent advances in machine learning, a new trend in database research
explores replacing programmed heuristics with learned ones~\citep{marcus2018towards,kipf2018learned, ortiz2018learning,marcus2018deep, btree, kraska2018case, mitzenmacher2018model, ma2018query}. 
Inspired by these results, this paper explores the natural question of synthesizing dataset-specific join search strategies using learning.
Assuming a given cost model and plan space, can we optimize the search over all possible join plans for a particular dataset? The hope is to learn tailored search strategies from the outcomes of previous planning instances that dramatically reduce search time for future planning. 

Our key insight is that join ordering has a deep algorithmic connection with 
Reinforcement Learning (RL)~\citep{sutton1998reinforcement}. Join ordering's sequential structure is the same problem structure that underpins RL. 
We exploit this algorithmic connection to embed RL deeply into a traditional query optimizer; anywhere an enumeration algorithm is used, a policy learned from an RL algorithm can just as easily be applied.
This insight enables us to achieve two key benefits. First, we can seamlessly integrate our solution into many optimizers with the classical System R architecture. Second, we exploit the nested structure of the problem to dramatically reduce the training cost, as compared to previous proposals for a ``learning optimizer''.  

To better understand the connection with RL, consider the classical ``bottom-up'' dynamic programming solution to join ordering.  The principle of optimality leads to an algorithm that incrementally builds a plan from optimal subplans of size two, size three, and so on.  Enumerated subplans are \emph{memoized} in a lookup table, which is consulted to construct a sequence of 1-step optimal decisions.  
Unfortunately, the space and time complexities of exact memoization can be prohibitive. Q-learning, an RL algorithm~\citep{sutton1998reinforcement}, relaxes the requirement of exact memoization. 
Instead, it formulates optimal planning as a prediction problem: given the costs of previously enumerated subplans, which 1-step decision is most likely optimal?
RL views the classic dynamic programming lookup table as a model---a data structure that \emph{summarizes} enumerated subplans and predicts the value of the next decision.
In concrete terms, Q-learning sets up a regression from the decision to join a particular pair of relations to the observed benefit of making that join on past data (i.e., impact on the final cost of the entire query plan).

To validate this insight, we built an RL-based optimizer \sys 
that optimizes select-project-join blocks and performs join ordering as well as physical operator selection. 
\sys observes the planning results of previously executed queries and trains an RL model to improve future search.  We implement three versions of \sys to illustrate the ease of integration into existing DBMSes: (1) A standalone version built on top of Apache Calcite~\citep{calcite}, (2) a version integrated with PostgreSQL~\citep{postgres}, and (3) a version integrated with SparkSQL~\citep{armbrust2015spark}.
Deploying \sys into existing production-grade systems (2) and (3) each required changes of less than 300 lines of code and training data could be collected through the normal operation of the DBMS with minimal overhead.

One might imagine that training such a model is extremely data-intensive.
While RL algorithms are indeed notoriously data-inefficient (typical RL settings, such as the Atari games~\citep{mnih2013playing}, require hundreds of thousands of training examples), we can exploit the optimal subplan structure specific to join optimization to collect an abundance of high-quality training data.  From a single query that passes through a native optimizer, not only are the final plan and its total cost collected as a training example, so are all of its subplans and, recursively, \textit{everything inside the exact memoization table}.  
For instance, planning an 18-relation join query in TPC-DS (Q64)
through a bushy optimizer can yield up to 600,000 training data points thanks to \sys's Q-learning formulation. 

We thoroughly study this approach on two workloads: Join Order Benchmark~\citep{leis2015good} and TPC-DS~\citep{tpcds}.%
\sys sees significant speedups in planning times (up to $>200\times$) relative to dynamic programming enumeration while essentially matching the execution times of optimal plans computed by the native enumeration-based optimizers.
These planning speedups allow for broadening the plan space to include bushy plans and Cartesian products.  In many cases, they lead to improved query execution times as well. 
\sys is particularly useful 
under
non-linear cost models such as memory limits or materialization.  On two simulated cost models with significant non-linearities, \sys improves on the plan quality of the next best heuristic over a set of 6 baselines by $1.7\times$ and $3\times$.  Thus, we show \sys approaches the optimization time efficiency of programmed heuristics \emph{and} the plan quality of optimal enumeration.

We are enthusiastic about the general trend of integrating 
learning techniques into database systems---not simply by black-box application of AI models to improve heuristics, but by the deep integration of algorithmic principles that span the two fields. Such an integration can facilitate new DBMS architectures that take advantage of all of the benefits of modern AI: learn from experience, adapt to new scenarios, and hedge against uncertainty. Our empirical results with \sys span across multiple systems, multiple cost models, and workloads. We show the benefits (and current limitations) of an RL approach to join ordering and physical operator selection. Understanding the relationships between RL and classical methods allowed us to achieve these results in a data-efficient way. We hope that \sys represents a step towards a future learning query optimizer.

\section{Background}

\begin{figure}[t]
    \centering
    \includegraphics[width=0.8\columnwidth]{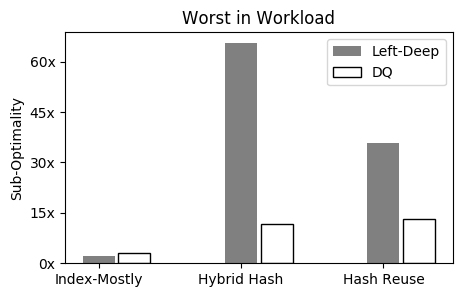}
    \caption{\small We consider 3 cost models for the Join Order Benchmark: (1) one with inexpensive index lookups, (2) one where the only physical operator is a hybrid hash join with limited memory, and (3) one that allows for the reuse of previously built hash tables. The figure plots the cost suboptimality w.r.t. optimal plans. 
    The classical left-deep dynamic program fails on the latter two scenarios. We propose a reinforcement learning based optimizer, \sys, which can adapt to a specific cost model given appropriate training data.
    \label{teaser}}
\end{figure}

The classic join ordering problem is, of course, NP-hard, and 
practical algorithms leverage heuristics to make the search for a good plan efficient.
The design and implementation of optimizer search heuristics are well-understood when the cost model is roughly linear, i.e., the cost of a join is linear in the size of its input relations.
This assumption underpins many classical techniques as well as recent work~\citep{selinger1979access,krishnamurthy1986optimization, trummer2017solving,neumann2018adaptive}.
However, many practical systems have relevant non-linearities in join costs.
For example, an intermediate result exceeding the available memory may trigger partitioning, or a relation may cross a size threshold that leads to a change in physical join implementation.

It is not difficult to construct reasonable scenarios where classical heuristics dramatically fail (Figure~\ref{teaser}).
Consider the query workload and dataset in the Join Order Benchmark~\citep{leis2015good}.
A popular heuristic from the original Selinger optimizer is to prune the search space to only include left-deep join orders.
Prior work showed that left-deep plans are extremely effective on this benchmark for cost models that prefer index joins~\citep{leis2015good}.
Experimentally, we found this to be true as well: the worst-case cost over the entire workload is only 2x higher than the true optimum (for an exponentially smaller search space).
However, when we simply change the cost model to be more non-linear, consisting of (1) hybrid hash join operators that spill partitions to disk when data size exceeds available memory, or (2) hash join operators that can re-use previously built hash tables, suddenly the left-deep heuristic is no longer a good idea---it is almost 50x more costly than the true optimum. 

These results illustrate that in a practical sense, the search problem is unforgiving: various heuristics have different weak spots where they fail by orders of magnitude relative to optimal.
For example, success on such atypical or non-linear cost models may require searching over ``bushy'' plans, not just left-deep ones.
With new hardware innovations~\citep{arulraj2017build} and a move towards serverless RDBMS architectures~\citep{aurora}, it is not unreasonable to expect a multitude of new query cost models that significantly differ from existing literature, which might require a complete redesign of standard pruning heuristics.
Ideally, instead of a fixed heuristic, we would want a strategy to guide the search space in a more data-driven way---tailoring the search to a specific database instance, query workload, and observed join costs.
This sets up the main premise of the paper: would it be possible to use data-driven machine learning methods to identify such a heuristic from data? 

\subsection{Example}
We focus on the classical problem of searching for a query plan made up of binary join operators and unary selections, projections, and access methods.
We will use the following database of three relations denoting employee salaries as a running example throughout the paper:
\[
\text{Emp}(id, name, rank) ~~ \text{Pos}(rank, title, code) ~~ \text{Sal}(code, amount)
\]
Consider the following join query:
\vspace{-0.7cm}
\begin{lstlisting}
SELECT *
  FROM Emp, Pos, Sal
 WHERE Emp.rank = Pos.rank 
   AND Pos.code = Sal.code
\end{lstlisting}
There are many possible orderings to execute this query. For example, one could execute the example query as $Emp \bowtie (Sal \bowtie Pos)$, or as $Sal \bowtie (Emp \bowtie Pos)$.

\subsection{Reinforcement Learning}
Bellman's ``Principle of Optimality'' and the characterization of dynamic programming is one of the most important results in computing~\citep{bellman2013dynamic}. In addition to forming the basis of relational query optimization, it has a deep connection to a class of stochastic processes called Markov Decision Processes (MDPs), which formalize a wide range of problems from path planning to scheduling. In an MDP model, an agent makes a sequence of decisions with the goal of optimizing a given objective (e.g., improve performance, accuracy). Each decision is dependent on the current state, and typically leads to  a new state.
The process is ``Markovian'' in the sense that the system's current state completely determines its future progression. 
Formally, an MDP consists of a five-tuple:
\[ \langle S, A, P(s,a), R(s,a), s_0 \rangle \]
where $S$ describes a set of states that the system can be in, $A$ describes the set of actions the agent can take, $s' \sim P(s,a)$ describes a probability distribution over new states given a current state and action, and  $s_0$ defines a distribution of initial states. $R(s,a)$ is the reward of taking action a in  state s. The reward measures the performance of the agent. 
The objective of an MDP is to find a decision policy $\pi: S \mapsto A$, a function that maps states to actions, with the maximum expected reward:
\begin{equation*}
\begin{aligned}
& \underset{\pi}{\argmax}
& & \mathbf{E}\left[\sum_{t=0}^{T-1} R(s_t,a_t)\right] \\
& \text{subject to}
& & s_{t+1} = P(s_t,a_t), a_t = \pi(s_t).
\end{aligned}
\end{equation*}
As with dynamic programming in combinatorial problems, most MDPs are  difficult to solve exactly. 
Note that the greedy solution, eagerly maximizing the reward at each step, might be suboptimal in the long run. Generally, analytical solutions to such problems scale  poorly in the time horizon.

Reinforcement learning (RL) is a class of  stochastic optimization techniques for MDPs~\citep{sutton1998reinforcement}.
An RL algorithm uses sampling, taking randomized sequences of decisions, to build a model that correlates decisions with improvements in the optimization objective (cumulative reward).
The extent to which the model is allowed to extrapolate depends on how the model is parameterized. 
One can parameterize the model with a table (i.e., exact parameterization) or one can use any function approximator (e.g., linear functions, nearest neighbors, or neural networks). 
Using a neural network in conjunction with RL, or Deep RL, is the key technique behind recent results like learning how to autonomously play Atari games~\citep{mnih2015human} and the game of Go~\citep{silver2016mastering}.

\subsection{Markov Model of Enumeration}
Now, we will review standard ``bottom-up'' join enumeration, and then, we will make the connection to a Markov Decision Process.
Every join query can be described as a query graph, where edges denote join conditions between tables and vertices denote tables.
Any dynamic programming join optimizer implementation needs to keep track of its progress: what has already been done in a particular subplan (which relations were already joined up) and what options remain (which relations--whether base or the result of joins--can still be ``joined in'' with the subplan under consideration). The query graph formalism allows us to represent this state. 

\begin{definition}[Query Graph]
A query graph $G$ is an undirected graph, 
where each relation $R$ is a vertex and each join predicate $\rho$ defines an edge between vertices. Let $\kappa_G$ denote the number of connected components of $G$.
\end{definition}

Making a decision to join two subplans corresponds to picking two vertices that are connected by an edge and merging them into a single vertex.
Let $G=(V,E)$ be a query graph.
Applying  a join $c=(v_i, v_j)$ to the graph $G$ defines a new graph with the following properties: (1) $v_i$ and $v_j$ are removed from $V$, (2) a new vertex $(v_i+v_j)$ is added to $V$, and (3) the edges of $(v_i+v_j)$ are the union of the edges incident to $v_i$ and $v_j$. 
Each join reduces the number of vertices by $1$. Each plan can be described as a sequence of such joins $c_1 \circ c_2 ...\circ c_{T}$ until $|V| = \kappa_G$. The above description embraces another System R heuristic: ``avoiding Cartesian products''. We can relax that heuristic by simply adding edges to $G$ at the start of the algorithm, to ensure it is fully connected.

Going back to our running example, suppose we start with a query graph consisting of the vertices $(Emp, Pos, Sal)$. Let the first join be $c_1 = (Emp, Pos)$; this leads to a query graph where the new vertices are $(Emp+Pos, Sal)$. Applying the only remaining possible join, we arrive at a single remaining vertex $Sal+(Emp+Pos)$ corresponding to the join plan $Sal \bowtie (Emp \bowtie Pos)$. 

The join optimization problem is to find the best possible join sequence---i.e., the best query plan. 
Also note that this model can be simply extended to capture physical operator selection as well. The set of allowed joins can be typed with an eligible join type, e.g., $c=(v_i, v_j, \textsf{HashJoin})$ or $c=(v_i, v_j, \textsf{IndexJoin})$.  We assume access to a cost model  $J(c) \mapsto \mathbb{R}_+$, i.e., a function that estimates the incremental cost of a particular join. 
\begin{problem}[Join Optimization Problem]
Let $G$ define a query graph and $J$ define a cost model. Find a sequence $c_1 \circ c_2 ...\circ c_{T}$ terminating in $|V| = \kappa_G$ to minimize:
\begin{equation*}
\begin{aligned}
& \min_{c_1,...,c_T}
& & \sum_{i=1}^T J(c_i) \\
& \text{subject to}
& & G_{i+1} =  c(G_i).
\end{aligned}
\end{equation*}
\label{joinopt}
\end{problem}
\vspace{-.25cm}
Note how this problem statement exactly defines an MDP (albeit by convention a minimization problem rather than maximization). $G$ is a representation of the \textbf{state}, $c$ is a representation of the \textbf{action}, the vertex merging process defines the state transition $P(G,c)$, and the reward function is the negative cost $-J$.
The output of an MDP is a function that maps a given query graph to the best next join.
Before proceeding, we summarize our notation in Table~\ref{table:notation}.

\subsection{Long Term Reward of a Join}

\begin{table}[t]\centering \small%
\ra{1.3}
\begin{tabular}{@{} l l @{}} \toprule
\textbf{\emph{Symbol}} & \textbf{\emph{Definition}}  \\ \midrule
   $G$  & A query graph.  This is a \emph{state} in the MDP.  \\ 
$c$ & A join. This is an \textit{action}. \\
$G'$   & The resultant query graph after applying a join. \\ 
$J(c)$   & A cost model that scores joins. \\ \bottomrule
\end{tabular}
\vspace{0.25em}
\caption{\small{Notation used throughout the paper.}\label{table:notation}}
\vspace{-0.8cm}
\end{table}

To introduce how RL gives us a new perspective on this classical database optimization problem, let us first examine the greedy solution.
A naive solution is to optimize each $c_i$ independently (also called Greedy Operator Optimization~\citep{neumann2018adaptive}). The algorithm proceeds as follows: (1) start with the query graph, (2) find the lowest cost join, (3) update the query graph and repeat until only one vertex is left. 

The greedy algorithm, of course, does not consider how local decisions might affect future costs. For illustration, consider our running example query with the following simple costs (assume a single join method with symmetric cost):
\[J(EP)= 100,~J(SP)= 90,~J((EP)S)= 10,~J((SP)E)= 50\]
The greedy solution would result in a cost of 140 (because it neglects the future effects of a decision), while the optimal solution has a cost of 110.
However, there is an upside: this greedy algorithm has a computational complexity of $O(|V|^3)$, despite the super-exponential search space.

The greedy solution is suboptimal because
the decision at each index fails to consider the long-term value of its action. One might have to sacrifice a short term benefit for a long term payoff.
Consider the optimization problem for a particular query graph $G$:
\begin{equation}
V(G) = \min_{c_1,...,c_T} \sum_{i=1}^T J(c_i)
\label{eq:main}
\end{equation}
In classical treatments of dynamic programming,
 this function is termed the \emph{value function}.  It is noted that optimal behavior over an entire decision horizon implies optimal behavior from any starting index $t>1$ as well, which is the basis for the idea of dynamic programming.
 Conditioned on the current join, we can write in the following form: 
\[
V(G) = \min_{c} Q(G,c)
\]
\[
Q(G,c) = J(c) + V(G')
\]
leading to the following recursive definition of the \emph{Q-function} (or cost-to-go function):
\begin{equation}
Q(G,c) = J(c) + \min_{c'} Q( G',c')
\label{eq:q}
\end{equation}
Intuitively, the Q-function describes the long-term value of each join: the cumulative cost if we act optimally for all subsequent joins after the current join decision. 
Knowing $Q$ is equivalent to solving the problem since local optimization $\min_{c'} Q(G',c')$ is sufficient to derive an optimal sequence of join decisions. 

If we revisit the greedy algorithm, and revise it hypothetically as follows: (1) start with the query graph, (2) find the lowest \emph{Q-value} join, (3) update the query graph and repeat, then this algorithm has the same computational complexity of $O(|V|^3)$ but is provably optimal.
To sketch out our solution, we will use Deep RL to approximate a global Q-function (one that holds for all query graphs in a workload), which gives us a polynomial-time algorithm for join optimization.


\subsection{Applying Reinforcement Learning}
\label{sec:apply-rl}
An important class of reinforcement learning algorithms, called Q-learning algorithms, allows us to approximate the Q-function from samples of data~\citep{sutton1998reinforcement}.
What if we could regress from features of $(G,c)$ to the future cumulative cost based on a small number of observations?
Practically, we can observe samples of decision sequences containing $(G,c, J(c), G')$ tuples, where $G$ is the query graph, $c$ is a particular join, $J(c)$ is the cost of the join, and $G'$ is the resultant graph.
Such a sequence can be extracted from any final join plan and by evaluating the cost model on the subplans.

Let's further assume we have a parameterized model for the Q-function, $Q_\theta$:
\[
Q_\theta(f_G,f_c) \approx Q(G,c)
\]
where $f_G$ is a \emph{feature vector} representing the query graph and $f_c$ is a feature vector representing a particular join. $\theta$ is the model parameters that represent this function and is randomly initialized at the start.
For each training tuple $i$, one can calculate the following label, or the ``estimated'' Q-value:
\[
y_i = J(c) + \min_{c'} Q_\theta(G',c')
\]
The $\{y_i\}$ can then be used as labels in a regression problem. If $Q$ were the true Q-function, then the following recurrence would hold:
\[
Q(G,c) = J(c) + \min_{c'} Q_\theta(G',c')
\]
So, the learning process, or \emph{Q-learning}, defines a loss  at each iteration:
\[
L(Q) = \sum_{i} \|y_i - Q_\theta(G,c)\|_2^2
\]
Then parameters of the Q-function can be optimized with gradient descent until convergence. 

RL yields two key benefits: (1) the search cost for a single query relative to traditional query optimization is radically reduced, since the algorithm has the time-complexity of greedy search, and (2) the parameterized model can potentially learn across queries that have ``similar'' but non-identical subplans.  
This is because the similarity between subplans are determined by the query graph and join featurizations, $f_G$ and $f_c$; thus if they are designed in a sufficiently expressive way, then the neural network can be trained to extrapolate the Q-function estimates to an entire workload.

The specific choice of Q-learning is important here (compared to other RL algorithms). First, it allows us to take advantage of optimal substructures during training and greatly reduce data needed. Second, compared to policy learning~\citep{marcus2018deep}, Q-learning outputs \emph{a score for each join that appears in any subplan} rather than simply selecting the best join. This is more amenable to deep integration with existing query optimizers, which have additional state like interesting orders and their own pruning of plans. Third, the scoring model allows for top-k planning rather than just getting the best plan. We note that the design of Q-learning variants is an active area of research in AI~\citep{hester2017deep, van2016deep}, so we opted for the simplicity of a Deep Q-learning approach and defer incorporation of advanced variants to future work.

\subsection{Reinforcement Learning vs. Supervised Learning}
Reinforcement Learning and Supervised Learning can seem very similar since the underlying inference methods in RL algorithms are often similar to those used in supervised learning and statistical estimation.
Here is how we justify our terminology.
In supervised learning, one has paired training examples with ground-truth labels (e.g., an image with a labeled object).
For join optimization, this would mean a dataset where the example is the current join graph and the label is the next best join decision from an oracle.
In the context of sequential planning, this problem setting is often called Imitation Learning~\citep{osa2018algorithmic}; where one imitates an oracle as best as possible.

As in~\citep{levine2018learning}, the term ``Reinforcement Learning'' refers to a class of empirical solutions to Markov Decision Process problems where we do \textit{not} have the ground-truth, optimal next steps; instead, learning is guided by numeric ``rewards'' for next steps. In the context of join optimization, these rewards are subplan costs. RL rewards may be provided by a real-world experiment, a simulation model, or some other oracular process. In our work below, we explore different reward functions including both real-world feedback (\secref{subsec:feedback}) and simulation via traditional plan cost estimation (\secref{sec:data-collection}).

RL purists may argue that access to any optimization oracle moves our formulation closer to supervised learning than classical RL.
We maintain this terminology because we see the pre-training procedure as a useful prior.
Rather than expensive, \emph{ab initio} learning from executions, we learn a useful (albeit imperfect) join optimization policy offline. 
This process bootstraps a more classical ``learning-by-doing'' RL process online that avoids executing grossly suboptimal query plans. 

There is additionally subtlety in the choice of algorithm. Most modern RL algorithms collect data episodically (execute an entire query plan and observe the final result). 
This makes sense in fields like robotics or autonomous driving where actions may not be reversible or decomposable. In query optimization,  every query consists of subplans (each of which is its own ``query''). Episodic data collection ignores this compositional structure.




\section{Optimizer Architecture}
\label{sec:optimizer-arch}
Selinger's optimizer design separated the problem of plan search from cost/selectivity estimation~\citep{selinger1979access}. This insight allowed independent innovation on each topic over the years. In our initial work, we follow this lead, and intentionally focus on learning a search strategy only. Even within the search problem, we focus narrowly on the classical select-project-join kernel. This too is traditional in the literature, going back to Selinger~\citep{selinger1979access} and continuing as recently as Neumann et al.'s very recent experimental work~\citep{neumann2018adaptive}. It is also particularly natural for illustrating the connection between dynamic programming and Deep RL and implications for query optimization. We intend for our approach to plug directly into a Selinger-based optimizer architecture like that of PostgreSQL, DB2 and many other systems.

In terms of system architecture, \sys can be simply integrated as a learning-based replacement for prior algorithms for searching a plan space. 
Like any non-exhaustive query optimization technique, our results are heuristic. The new concerns raised by our approach have to do with limitations of training, including overfitting and avoiding high-variance plans. 
We use this section to describe the extensibility of our approach and what design choices the user has at her disposal.

\def \treeA {\tikz[inner sep=1pt, baseline=(T3.north)] {
                  \node (T1) at (0,0) {$T_1$};
                  \node (T2) at (1,0) {$T_2$};
                  \node (INLJ) at (0.5,0.5) {\small{\sffamily IndexJoin}};

                  \node (T3) at (1.5,0.5) {$T_3$};
                  \node (SMJ) at (1,1) {\small{\sffamily HashJoin}};

                  \node (T4) at (2,1) {$T_4$};
                  \node (HJ) at (1.5,1.5) {\small{\sffamily HashJoin}};

                  \draw (T1)--(INLJ)--(T2);
                  \draw (INLJ)--(SMJ)--(T3);
                  \draw (SMJ)--(HJ)--(T4);}}

\def \treeB {\tikz[inner sep=1pt, baseline=(T3.south)] {
                  \node (T1) at (0,0) {$T_1$};
                  \node (T2) at (1,0) {$T_2$};
                  \node (INLJ) at (0.5,0.5) {\small{\sffamily IndexJoin}};

                  \node (T3) at (1.5,0.5) {$T_3$};
                  \node (SMJ) at (1,1) {\small{\sffamily HashJoin}};

                  \draw (T1)--(INLJ)--(T2);
                  \draw (INLJ)--(SMJ)--(T3);}}

\def\treeC{\tikz[inner sep=1pt, baseline=(T1.north)] {
                  \node (T1) at (0,0) {$T_1$};
                  \node (T2) at (1,0) {$T_2$};
                  \node (INLJ) at (0.5,0.5) {\small{\sffamily IndexJoin}};
                  \draw (T1)--(INLJ)--(T2);}}%

\begin{figure*}[!h]
        \begin{tikzpicture}[inner sep=1pt,text centered]
          \node (TA) at (0,0) {$(~~~\{$ \treeA , \treeB , \treeC   $\}$,~~~ $\{T1,\cdots,T4\}$;~~~ $\quad V^*)$};
          \node[left = of TA] (orig) {\treeA};
          \draw [double,->, >=stealth] (orig)--(TA);
          \node[text width=2.5cm] (plan) at (-7,-1.5)  {\footnotesize Plan from Native Optimizer};
          \node (plan) at (-1.3,-1.5)  {\footnotesize Optimal Sub-plans};
          \node[text width=1.5cm] (plan) at (2.75, -1.5)  {\footnotesize Relations to Join};
          \node[text width=1.5cm] (plan) at (4.35, -1.5)  {\footnotesize Optimal Cost};
          
          \node[rectangle, left = of orig, rounded corners=5, text width=2.0cm, draw] (opt) {\small Native Optimizer};
        \draw[double,->, >=stealth] (opt) -- (orig);
        \end{tikzpicture}
        \vspace{-.3cm}
      \caption{\small Training data collection is efficient (\secref{sec:data-collection}).  Here, by leveraging the principle of
      optimality, three training examples are emitted from a single
      plan produced by a native optimizer.  These examples share the same long-term cost and relations to join (i.e., making these local decisions eventually leads to joining $\{T1, \cdots, T4\}$ with optimal cumulative cost $V^*$).}
      \label{fig:data-collection}
\end{figure*}
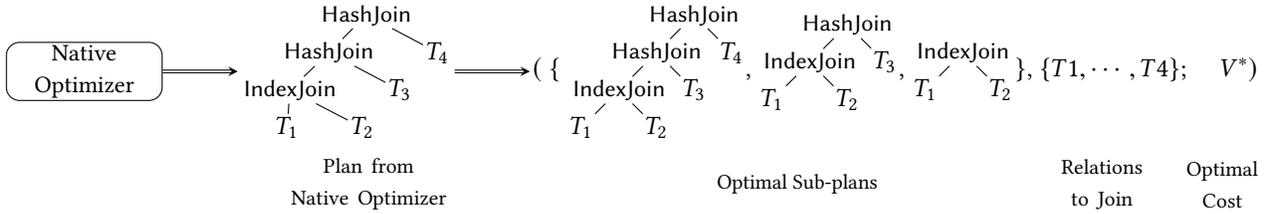

\subsection{Overview}
Now, we describe what kind of training data is necessary to learn a Q-function. In supervised regression, we collect data of the form \texttt{(feature, values)}. The learned function maps from feature to values. One can think of this as a \emph{stateless} prediction, where the underlying prediction problem does not depend on some underlying process state. On the other hand, in the Q-learning setting, there is state. So we have to collect training data of the form \texttt{(state, decision, new state, cost)}. Therefore, a training dataset has the following format (in Java notation):
\vspace{-0.3cm}
\begin{lstlisting}[aboveskip=0pt,language=java]
List<Graph, Join, Graph', Cost> dataset
\end{lstlisting}

In many cases like robotics or game-playing, RL is used in a live setting where the model is trained on-the-fly based on concrete moves chosen by the policy and measured in practice. 
Q-learning is known as an ``off-policy'' RL method. 
This means that its training is independent of the data collection process and can be suboptimal---as long as the training data sufficiently covers the decisions to be made.

\subsection{Architecture and API}
\sys collects training data sampled from a cost model and a native optimizer. It builds a model which improves future planning instances.
\sys makes relatively minimal assumptions about the structure of the optimizer. Below are the API hooks that it requires implemented.

\vspace{.5em} \noindent \emph{Workload Generation.} A function that returns a list of training queries of interest. \sys requires a relevant workload for training. In our experiments, we show that this workload can be taken from query templates or sampled from the database schema.
\vspace{-0.3cm}
\begin{lstlisting}[aboveskip=0pt,language=java]
sample(): List<Queries> 
\end{lstlisting}
\vspace{0.5em}

\noindent \emph{Cost Sampling.} A function that given a query returns a list of join actions and their resultant costs. \sys requires the system to have its own optimizer to generate training data. This means generating feasible join plans and their associated costs. Our experiments evaluate integration with deterministic enumeration, randomized, and heuristic algorithms.
\vspace{-0.3cm}
\begin{lstlisting}[aboveskip=0pt,language=java]
train(query): List<Graph,Join,Graph',Cost> 
\end{lstlisting}
\vspace{0.5em}

\noindent \emph{Predicate Selectivity Estimation.} A function that returns the selectivity of a particular single table predicate. \sys leverages the optimizer's own selectivity estimate for featurization (\secref{subsec:featurization}). 
\vspace{-0.3cm}
\begin{lstlisting}[aboveskip=0pt,language=java]
selectivity(predicate): Double
\end{lstlisting}
\vspace{0.5em}

In our evaluation (\secref{sec:eval}), we will vary these exposed hooks to experiment with different implementations for each (e.g., comparing training on highly relevant data from a desired workload vs. randomly sampling join queries directly from the schema).

\subsection{Efficient Training Data Generation}
\label{sec:data-collection}
Training data generation may seem onerous, but in fact, useful data is \emph{automatically generated} as a consequence of running classical planning algorithms. 
For each join decision that the optimizer makes, we can get the incremental cost of the join.
Suppose, we run a classical bushy dynamic programming algorithm to optimize a k-way join, we not only get a final plan 
but also an optimal plan for every single subplan enumerated along the way.
Each query generates an optimal query plan for all of the subplans that compose it, as well as observations of suboptimal plans that did not make the cut.
This means that a single query generates a large amount of training examples.
Figure \ref{fig:data-collection} shows how the principle of optimality helps enhance a training dataset. 

This data collection scheme differs from that of several popular 
RL algorithms such as PPO and Policy Gradients~\citep{schulman2017proximal} (and used in ~\citep{marcus2018deep}). 
These algorithms train their models ``episodically'', where they apply an entire sequence of decisions and observe the final cumulative reward.
An analogy would be a graph search algorithm that does not backtrack but resets to the starting node and tries the whole search again.
While  general, this scheme not suited for the structure of join optimization, where an optimal plan is composed of optimal substructures.
Q-learning, an algorithm that does not rely on episodic data and can learn from offline data consisting of a hierarchy of optimal subplans, is a better fit for join optimization. 


\begin{figure*}[!htp]
\captionsetup[subfigure]{justification=centering}
\begin{subfigure}[b]{.23\textwidth}
\raisebox{10mm}
\centering
\small
\begin{verbatim}
SELECT *
  FROM Emp, Pos, Sal
 WHERE Emp.rank 
       = Pos.rank 
   AND Pos.code 
       = Sal.code
\end{verbatim}
\caption{Example query}
\end{subfigure}\hspace{\fill}
\begin{subfigure}[b]{.23\textwidth}
\centering
\small
\begin{equation*}
\begin{split} 
A_G = [ & \text{E.id, E.name, E.rank,} \\ 
& \text{P.rank, P.title, P.code,} \\
& \text{S.code, S.amount} ] \\
= [ & 1\ 1\ 1\ 1\ 1\ 1\ 1\ 1 ]
\end{split} 
\end{equation*}
    \vspace{-.2cm}
\caption{Query graph featurization \label{fig:query-graph-feat}}
\end{subfigure}\hspace{\fill}
\begin{subfigure}[b]{.23\textwidth}
\centering
\small
\begin{equation*}
\begin{split} 
A_L &= [ \text{E.id, E.name, E.rank} ] \\
&=  [ 1\ 1\ 1\ 0\ 0\ 0\ 0\ 0 ] \\
A_R &= [ \text{P.rank, P.title, P.code}  ] \\
&=  [ 0\ 0\ 0\ 1\ 1\ 1\ 0\ 0 ]
\end{split} 
\end{equation*}
    \vspace{-.2cm}
\caption{Features of $E \bowtie P$}
\end{subfigure}\hspace{\fill}
\begin{subfigure}[b]{.23\textwidth}
\centering
\small
\begin{equation*}
\begin{split} 
A_L = [ & \text{E.id, E.name, E.rank,} \\
& \text{P.rank, P.title, P.code} ] \\
=  [ & 1\ 1\ 1\ 1\ 1\ 1\ 0\ 0 ] \\
A_R = [ & \text{S.code, S.amount}  ] \\
=  [& 0\ 0\ 0\ 0\ 0\ 0\ 1\ 1 ]
\end{split} 
\end{equation*}
    \vspace{-.2cm}
\caption{Features of $(E \bowtie P) \bowtie S$}
\end{subfigure}
\vspace{-.2cm}
\caption{\small{
{\bf A query and its corresponding featurizations (\secref{subsec:featurization}).} One-hot vectors encode the visible attributes in the query graph ($A_G$), the left side of a join ($A_L$), and the right side ($A_R$).   Such encoding allows for featurizing both the query graph and a particular join.  A partial join and a full join are shown. The example query covers all relations in the schema, so $A_G = A$. \label{fig:feat}
}}
\end{figure*}

In our experiments, we bootstrap planning with a bushy dynamic program until the number of relations in the join exceeds 10 relations.  Then, the data generation algorithm switches to a greedy scheme for efficiency for the last $K-10$ joins.
Ironically, the data collected from such an optimizer might be ``too good'' (or too conservative) because it does not measure or learn from a diverse enough space of (costly, hence risky) subplans.
If the training data only consisted of optimal sub-plans, then the learned Q-function may not accurately learn the downside of poor subplans. Likewise, if purely random plans are sampled, the model might not see very many instances of good plans. 
To encourage more ``exploration'', during data collection noise can be injected into the optimizer to force it to enumerate more diverse subplans.
We control this via a parameter $\epsilon$, 
the probability of picking a random join as opposed to a join with the lowest cost.  
As the algorithm enumerates subplans, if  $\textsf{rand()} < \epsilon$ then a random (valid) join is chosen on the current query graph; otherwise it proceeds with the lowest-cost join as usual.
This is an established technique to address such ``covariate shift'', a phenomenon extensively studied in prior work~\citep{laskey2017dart}.


\begin{figure}[!t]
\captionsetup[subfigure]{justification=centering}
\begin{subfigure}[b]{.45\columnwidth}
\small
Query:
\vspace{-.1cm}
\begin{verbatim}
<example query>
  AND Emp.id > 200
\end{verbatim}
\begin{equation*}
\begin{split} 
\text{Selectivity}(\text{Emp.id$>$200}) = 0.2
\end{split} 
\end{equation*}
\vspace{.03cm}
\begin{equation*}
\begin{split} 
f_G = A_G =&\ [ \text{E.id, E.name}, \cdots ] \\
=& \ [ 1\ 1\ 1\ 1\ 1\ 1\ 1\ 1 ] \\
\rightarrow& \ [ .2\ 1\ 1\ 1\ 1\ 1\ 1\ 1 ]
\end{split} 
\end{equation*}
\caption{Selectivity scaling in query graph features \label{fig:feat-sel-scaling}}
    \vspace{.2cm}
\end{subfigure}\hspace{\fill}%
\begin{subfigure}[b]{.45\columnwidth}
\small
Query:
\vspace{-.1cm}
\begin{verbatim}
<example query>
\end{verbatim}
\vspace{.3cm}
\begin{equation*}
\begin{split} 
\text{fe} & \text{at\_vec}(\textsf{IndexJoin}(E \bowtie P)) \\
& = A_L \oplus A_R \oplus [1\ 0]
\end{split} 
\end{equation*}
\vspace{.1cm}
\begin{equation*}
\begin{split} 
\text{fe} & \text{at\_vec}(\textsf{HashJoin}(E \bowtie P)) \\
& = A_L \oplus A_R \oplus [0 \ 1]
\end{split} 
\end{equation*}
\caption{Concatenation of physical operators in join features \label{fig:feat-physical-op}}
    \vspace{.2cm}
\end{subfigure}
\vspace{-.3cm}
\caption{\small{{\bf Accounting for selections and physical operators.}  Simple changes to the basic form of featurization are needed to support selections (left) and physical operators (right). For example, assuming a system that chooses between only \textsf{IndexJoin} and \textsf{HashJoin}, a 2-dimensional one-hot vector is concatenated to each join feature vector. Discussion in \secref{subsec:featurization}.}}
\vspace{-0.4cm}
\end{figure}

\section{Realizing the Q-learning Model}
Next, we present the mechanics of actually training and operating a Q-learning model.

\subsection{Featurizing the Join Decision}
\label{subsec:featurization}
Before we get into the details, we will give a brief motivation of how we should think about featurization in a problem like this.
The features should be sufficiently rich that they capture all relevant information to predict the future cumulative cost of a join decision.
This requires knowing what the overall query is requesting, the tables on the left side of the proposed join, and the tables on the right side of the proposed join.
It also requires knowing how single table predicates affect cardinalities on either side of the join.

\vspace{0.5em} \noindent \textbf{Participating Relations: } The overall intuition is to use each column name as a feature, because it identifies the distribution of that column.  The first step is to construct a set of features to represent which attributes are participating in the query and in the particular join. Let $A$ be the set of all attributes in the database (e.g., $\{Emp.id, Pos.rank,...,Sal.code,Sal.amount\}$). Each relation $rel$ (including intermediate join results) has a set of \emph{visible attributes}, $A_{rel} \subseteq A$, the attributes present in the output. Similarly, every query graph $G$ can be represented by its visible attributes $A_G \subseteq A$. Each join is a tuple of two relations $(L,R)$ and we can get their visible attributes $A_L$ and $A_R$. Each of the attribute sets $A_G, A_L, A_R$ can then be represented with a \emph{binary 1-hot encoding}: a value $1$ in a slot indicates that particular attribute is present, otherwise $0$ represents its absence.  
Using $\oplus$ to denote concatenation, we obtain the query graph features,  $f_{G} = A_{G}$, and the join decision features, $f_{c} = A_{L} \oplus A_{R}$, and, finally, the overall featurization for a particular $(G,c)$ tuple is simply $f_{G} \oplus f_c$.  Figure \ref{fig:feat} illustrates the featurization of our example query.


\vspace{0.5em} \noindent \textbf{Selections: } Selections can change said distribution, i.e., \textsf{(col, sel-pred)} is different than \textsf{(col, TRUE)}.
To handle single table predicates in the query, we have to tweak the feature representation. As with most classical optimizers, we assume that the optimizer eagerly applies selections and projections to each relation. 
Next, we leverage the table statistics present in most RDBMS. For each selection $\sigma$ in a query we can obtain the selectivity $\delta_{\sigma}$, which estimates the fraction of tuples present after applying the selection.\footnote{We consider selectivity estimation out of scope for this paper. See discussion in \secref{sec:optimizer-arch} and \secref{sec:relatedwork}.}
To account for selections in featurization, we simply scale the slot in $f_G$  that the relation and attribute $\sigma$ corresponds to, by $\delta_r$. 
For instance, if selection $\text{Emp.id} > 200$ is estimated to have a selectivity of $0.2$, then the $\text{Emp.id}$ slot in $f_G$ would be changed to $0.2$.
Figure \ref{fig:feat-sel-scaling} pictorially illustrates this scaling.

\vspace{0.5em} \noindent \textbf{Physical Operators: } The next piece is to featurize the choice of physical operator. This is straightforward: we add another one-hot vector that indicates from a fixed set of implementations the type of join used (Figure~\ref{fig:feat-physical-op}).


\vspace{0.5em} \noindent \textbf{Extensibility: } 
In this paper, we focus only on the basic form of featurization described above and study foreign key equality joins.\footnote{This is due to our evaluation workloads containing only such joins. \secref{sec:extensions} discusses how \sys could be applied to more general join types.}  An ablation study as part of our evaluation (Table~\ref{table:feat-ablation}) shows that the pieces we settled on all contribute to good performance.  That said, there is no architectural limitation in \sys that prevents it from utilizing other features.  Any property believed to be relevant to join cost prediction can be added to our featurization scheme.
For example, we can add an additional binary vector  $f_{ind}$ to indicate which attributes have indexes built.  Likewise, physical properties like sort-orders can be handled by indicating which attributes are sorted in an operator's output.  Hardware environment variables (e.g., available memory) can be added as scalars if deemed as important factors in determining the final best plan. 
Lastly, more complex join conditions such as inequality conditions can also be handled (\secref{sec:extensions}).



\subsection{Model Training}
 \sys uses a multi-layer perceptron (MLP) neural network to represent the Q-function. 
It takes as input the final featurization for a $(G,c)$ pair, $f_G \oplus f_c$.
Empirically, we found that a two-layer MLP offered the best performance under a modest training time constraint ($< 10$ minutes). 
The model is trained with a standard stochastic gradient descent (SGD) algorithm.

\subsection{Execution after Training}
\label{sec:inference}
After training, we obtain a parameterized estimate of the Q-function, $Q_\theta(f_G,f_c)$. For execution, we simply go back to the standard algorithm as in the greedy method but instead of using the local costs, we use the learned Q-function: (1) start with the query graph, (2) featurize each join, (3) find the join with the lowest \emph{estimated Q-value} (i.e., output from the neural net), (4) update the query graph and repeat.

This algorithm has the time-complexity of greedy enumeration except in greedy, the cost model is evaluated at each iteration, and in our method, a neural network is evaluated.  One pleasant consequence is that \sys exploits the abundant vectorization opportunities in numerical computation.  In each iteration, instead of invoking the neural net sequentially on each join's feature vector, \sys \emph{batches} all candidate joins (of this iteration) together, and invokes the neural net once on the batch.  Modern CPUs, GPUs, and specialized accelerators (e.g., TPUs~\citep{jouppi2017datacenter}) all offer optimized instructions for such single-instruction multiple-data (SIMD) workloads. 
The batching optimization amortizes each invocation's fixed overheads and has the most impact on large joins.


\section{Feedback From Execution}
\label{subsec:feedback}

\begin{table*}[th!]\centering \small%
\begin{tabular}{@{} l l l l l l l l l l l l l l @{}} \toprule
{\bf Optimizer}  && \multicolumn{3}{c}{\bf Cost Model 1}  & & \multicolumn{3}{c}{\bf Cost Model 2} & & \multicolumn{3}{c}{\bf Cost Model 3}\\
  && {Min}  & {Mean}  & {Max} && {Min}  & {Mean}  & {Max} && {Min}  & {Mean}  & {Max}\\ \midrule
QuickPick ({\bf QP}) && 1.0  & 23.87 & 405.04 && 7.43  & 51.84 & 416.18 && 1.43  & 16.74 & 211.13 \\
IK-KBZ ({\bf KBZ}) && 1.0  & 3.45  & 36.78 && 5.21  & 29.61  & 106.34  && 2.21  & 14.61  & 96.14 \\
Right-deep ({\bf RD}) && 4.70 & 53.25 & 683.35 && 1.93 & 8.21 & 89.15 && 1.83 & 5.25 & 69.15 \\
Left-deep ({\bf LD}) && 1.0  & 1.08   & 2.14 && 1.75  & 7.31   & 65.45  && 1.35  & 4.21   & 35.91 \\
Zig-zag ({\bf ZZ}) && 1.0  & 1.07   & 1.87 && 1.0  & 5.07   & 43.16  && 1.0  & 3.41   & 23.13 \\
Exhaustive ({\bf EX}) && 1.0  & 1.0   & 1.0 && 1.0  & 1.0   & 1.0  && 1.0  & 1.0   & 1.0 \\
\sys  && 1.0  & 1.32   & 3.11 && 1.0  & 1.68   & 11.64 && 1.0  & 1.91   & 13.14 \\  \bottomrule
\end{tabular}
\vspace{0.25em}
\caption{\small{\sys is robust and competitive under all three cost models
(\secref{subsec:standalone}).  Plan costs are relative to optimal plans produced by exhaustive enumeration, i.e., $cost_{algo}/cost_{\textbf{EX}}$.  Statistics are calculated across the entire Join Order Benchmark.} \label{table:standalone-combined}}
\vspace{-0.6cm}
\end{table*}

We have described how \sys learns from sampling the cost model native to a query optimizer.  
However, it is well-known that a cost model (costs) may fail to correlate with reality (runtimes), due to poor cardinality estimates or unrealistic rules used in estimation.  To correct these errors, the database community has seen proposals of leveraging feedback from execution~\citep{chaudhuri2008pay, markl2003leo}.  We can perform an analogous operation on learned Q-functions.  Readers might be familiar with the concept of fine-tuning in the deep learning literature~\citep{yosinski2014transferable}, where a network is trained on one dataset and ``transferred'' to another with minimal re-training.  \sys can optionally apply this technique to re-train itself on real execution runtimes to correlate better with the operating environment.

\subsection{Fine-tuning \sys}
Fine-tuning \sys consists of two steps: pre-training as usual and re-training.  First, \sys is pre-trained to convergence on samples from the optimizer's cost model; these are inexpensive to collect compared to real execution.  Next, the weights of the first two layers of the neural network are frozen, and the output layer's weights are re-initialized randomly.  Re-training is then started on samples of real execution runtimes, which would only change the output layer's weights.  

Intuitively, the process can be thought of as first using the cost model to learn relevant features about the general structure of subplans (e.g., ``which relations are generally beneficial to join?'').  The re-trained output layer then projects the effect of these features onto real runtimes.
Due to its inexpensive nature, partial re-training is a common strategy applied in many machine learning applications.

\subsection{Collecting Execution Data}

For fine-tuning, we collect a list of real-execution data, \texttt{(Graph, Join, Graph', OpTime)}, where instead of the cost of the join, the real runtime attributed to the particular join operator is recorded.
Per-operator runtimes can be collected by instrumenting the underlying system, or using the system's native analysis functionality (e.g., \textsf{EXPLAIN ANALYZE} in Postgres).


\section{Evaluation}
\label{sec:eval}
We extensively evaluate \sys 
to investigate the following major questions:
\begin{itemize}
    \item How effective is \sys in producing plans, how good are they, and under what conditions (\secref{sec:eval-cm1}, \secref{sec:eval-cm2}, \secref{sec:eval-cm3})?
    \item How efficient is \sys at producing plans, in terms of runtimes and required data (\secref{eval:plan-latency}, \secref{eval:data-quantity}, \secref{eval:data-relevance})?
    \item Do \sys's techniques apply to real-world scenarios, systems, and workloads (\secref{eval:real-systems}, \secref{sec:eval-feedback})?
\end{itemize}
To address the first two questions, we run experiments on standalone \sys.  The last question is evaluated with end-to-end experiments on \sys-integrated Postgres and SparkSQL.

\subsection{Standalone Optimization Experiments}
\label{subsec:standalone}
We implemented \sys and a wide variety of optimizer search techniques previously benchmarked in Leis et al.~\citep{leis2015good} in a standalone Java query optimizer harness.
Apache Calcite is used for parsing SQL and representing the SQL AST. 
We first evaluate standalone \sys and other optimizers for final plan costs; unless otherwise noted, exploration (\secref{sec:data-collection}) and real-execution feedback (\secref{subsec:feedback}) are turned off.
We use the Join Order Benchmark (JOB)~\citep{leis2015good}, which is derived from the real IMDB dataset (3.6GB in size; 21 tables).
The largest table has 36 million rows.
The benchmark contains 33 templates and 113 queries in total.  The joins have between 4 and 15 relations, with an average of 8 relations per query.

We revisit a motivating claim from earlier: heuristics are well-understood when the cost model is linear but non-linearities can lead to significant suboptimality. 
The experiments intend to illustrate that \sys offers a form of \emph{robustness to cost model}, meaning, that it prioritizes plans tailored to the structure of the cost model, workload, and physical design---even when these plans are bushy.

We consider 3 cost models: CM1 is a model for a main-memory database; CM2 additionally considers limited memory hash joins where after a threshold the costs of spilling partitions to disk are considered; CM3 additionally considers the re-use of already-built hash tables during upstream operators. We compare with the following baselines: QuickPick-1000 (\textbf{QP})~\citep{waas2000join} selects the best of 1000 random join plans; IK-KBZ (\textbf{KBZ})~\citep{krishnamurthy1986optimization} is a polynomial-time heuristic that decomposes the query graph into chains and orders them; dynamic programs Right-deep (\textbf{RD}), Left-deep (\textbf{LD}), Zig-zag (\textbf{ZZ})~\citep{ziane1993parallel}, and Exhaustive (\textbf{EX}) exhaustively enumerate join plans with the indicated plan shapes.  
Details of the setup are listed in Appendix~\secref{appendix:cost-details}.  

Results of this set of experiments are shown in Table~\ref{table:standalone-combined}.

\subsubsection{Cost Model 1}
\label{sec:eval-cm1}
Our results on CM1 reproduce the conclusions of Leis et al.~\citep{leis2015good}, where left-deep plans are generally good (utilize indexes well) and there is little need for zigzag or exhaustive enumeration. \sys is competitive with these optimal solutions without \emph{a priori} knowledge of the index structure. In fact, \sys significantly outperforms the other heuristic solutions \textbf{KBZ} and \textbf{QP}.
While it is true that \textbf{KBZ} also restricts its search to left-deep plans, it is suboptimal for cyclic join graphs---its performance is hindered since almost all JOB queries contain cycles. 
We found that \textbf{QP} struggles with the physical operator selection, and a significant number of random samples are required to find a narrow set of good plans (ones the use indexes effectively).


Unsurprisingly, these results show that \sys, a learning-based solution, reasonably matches performance on cases where good heuristics exist.
On average \sys is within 22\% of the \textbf{LD} solution and in the worst case only 1.45$\times$ worse.

\subsubsection{Cost Model 2}
\label{sec:eval-cm2}
By simply changing to a different, yet realistic, cost model, we can force the left-deep heuristics to perform poorly.
CM2 accounts for disk usage in hybrid hash joins.
In this cost model, none of the heuristics match the exhaustive search over the entire workload.
Since the costs are largely symmetric for small relation sizes, there is little benefit to either left-deep or right-deep pruning.
Similarly zig-zag trees are only slightly better, and the heuristic methods fail by orders-of-magnitude on their worst queries.


\sys still comes close to the quality of exhaustive enumeration ($1.68\times$ on average). It does not perform as well as in CM1 (with its worst query about $12\times$ the optimal cost) but is still significantly better than the alternatives. Results on CM2 suggest that as memory becomes more limited, heuristics begin to diverge more from the optimal solution. We explored this phenomenon further and report results in Table~\ref{table:mem-limit}.


\begin{table}[t!]\centering \small%
\begin{tabular}{@{} l l l l l @{}} \toprule
    & {$M=10^8$}  & {$M=10^6$}  & {$M=10^4$} & {$M=10^2$}  \\ \midrule
{\bf KBZ} & 1.0  & 3.31  & 30.64 & 41.64  \\ 
{\bf LD}  & 1.0  & 1.09   & 6.45 & 6.72  \\ 
{\bf EX}  & 1.0  & 1.0   & 1.0 & 1.0  \\
\sys & 1.04  & 1.42   & 1.64 & 1.56 \\ 
 \bottomrule
\end{tabular}
\vspace{0.25em}
\caption{\small{Cost Model 2: mean relative cost vs. memory limit (number of tuples in memory).} \label{table:mem-limit}}
\vspace{-0.6cm}
\end{table}

\subsubsection{Cost Model 3}
\label{sec:eval-cm3}
Finally, we illustrate results on CM3 that allows for the reuse of hash tables.
Right-deep plans are no longer inefficient in this model as they facilitate reuse of the hash table (note right and left are simply conventions and there is nothing important about the labels). The challenge is that now plans have to contain a mix of left-deep and right-deep structures.
Zig-zag tree pruning heuristic was exactly designed for cases like this.
Surprisingly, \sys is significantly ($1.7\times$ on average and in the worst) better than zig-zag enumeration.
We observed that bushy plans were necessary in a small number of queries and \sys found such lower-cost solutions.


In summary, 
results in Table~\ref{table:standalone-combined} show that \sys is robust against different cost model regimes, since it learns to adapt to the workload at hand.

\subsubsection{Planning Latency}
\label{eval:plan-latency}
Next, we report the planning (optimization) time of \sys and several other optimizers across the entire 113 JOB queries.  
The same model in \sys is used to plan all queries.
Implementations are written in Java, single-threaded\footnote{To ensure fairness, for \sys we configure the underlying linear algebra library to use 1 thread.  No GPU is used.},
and reasonably optimized at the algorithmic level (e.g., QuickPick would short-circuit a partial plan already estimated to be more costly than the current best plan)---but no significant efforts are spent on low-level engineering. 
Hence, the relative magnitudes are more meaningful than the absolute values. 
Experiments were run on an AWS EC2 c5.9xlarge instance with a 3.0GHz 
CPU and 72GB memory.



\begin{figure}
    \centering
    \includegraphics[width=0.9\columnwidth,keepaspectratio]{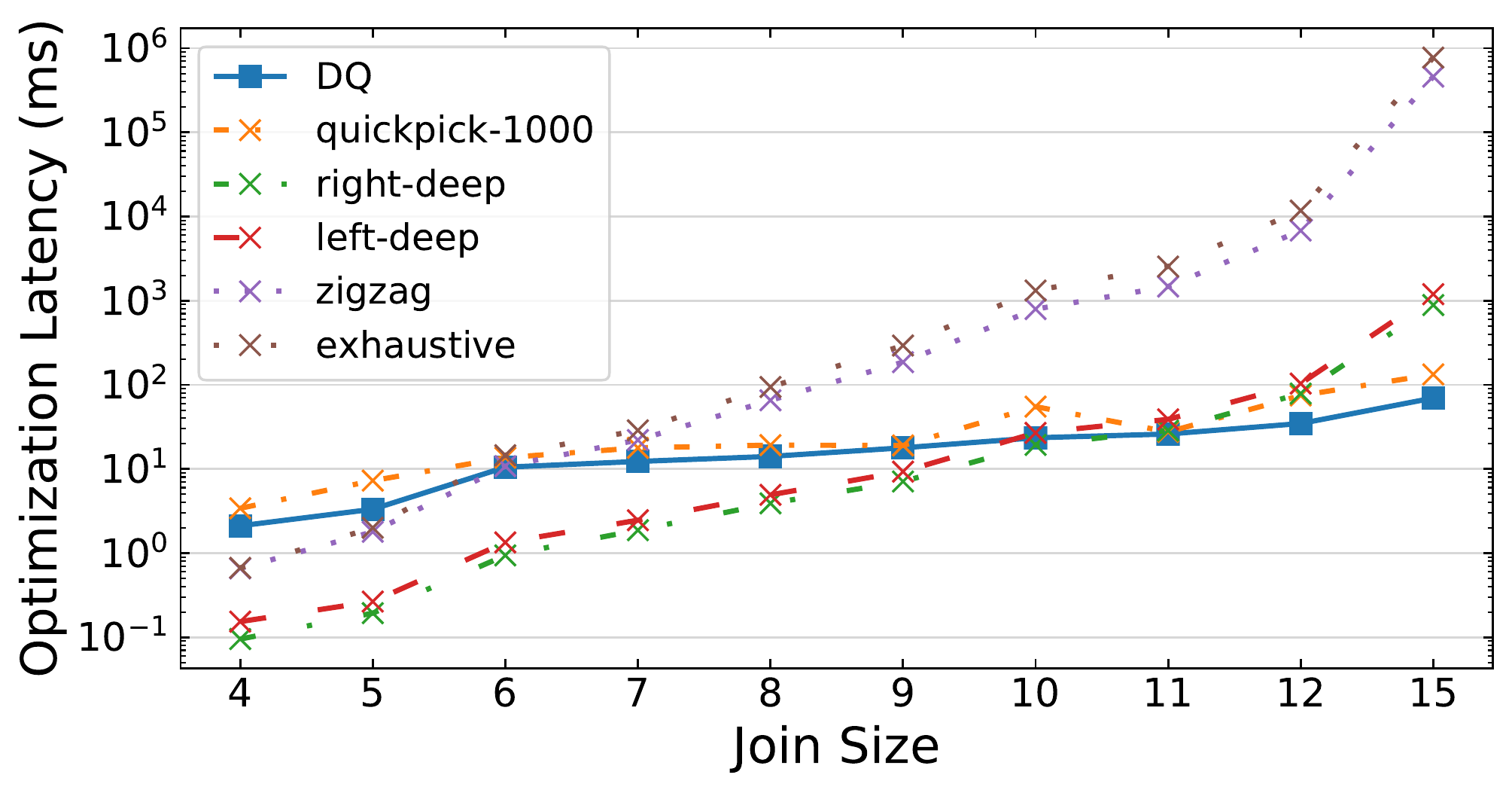}
\vspace{-0.5cm}
    \caption{\small{
    Optimization latency (log-scale) on all JOB queries grouped by number of relations in each query (\secref{eval:plan-latency}). A total of 5 trials are run; standard deviations are negligible hence omitted.%
    } \label{exp:planning-latency}}
    \vspace{-.3cm}
\end{figure}

Figure ~\ref{exp:planning-latency} reports the runtimes grouped by number of relations. In the small-join regime, \sys's overheads are attributed interfacing with a JVM-based deep learning library, \textsf{DL4J} (creating and filling the featurization buffers; JNI overheads due to native CPU backend execution).  These could have been optimized away by targeting a non-JVM engine and/or GPUs, but we note that when the number of joins is small, exhaustive enumeration would be the ideal choice.

In the large-join regime, \sys achieves drastic speedups: for the largest joins \sys runs up to 10,000$\times$ faster than exhaustive enumeration and $>10\times$ than left-deep.  
\sys upper-bounds the number of neural net invocations by the number of relations in a query, and additionally benefits from the batching optimization (\secref{sec:inference}). 
We believe this is a profound performance argument for a learned optimizer---it would have an even more unfair advantage when applied to larger queries or executed on specialized accelerators~\citep{jouppi2017datacenter}.


\subsubsection{Quantity of Training Data}
\label{eval:data-quantity}
\begin{figure}
    \centering
    \includegraphics[width=0.9\columnwidth,keepaspectratio]{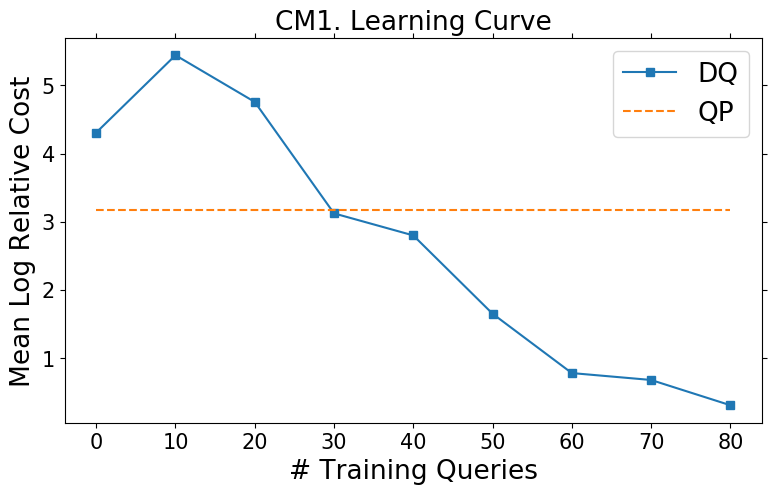}
\vspace{-0.4cm}
    \caption{\small{Mean relative cost (in log-scale) as a function of the number of training queries seen by \sys. We include QuickPick-1000 as a baseline.  Cost Model 1 is used.} \label{exp:plot2}}
\vspace{-0.1cm}
\end{figure}

How much training data does \sys need to become effective?  To study this, we vary the number of training queries given to \sys and plot the mean relative cost using the cross validation technique described before. Figure \ref{exp:plot2} shows the relationship. \sys requires about 60-80 training queries to become competitive and about 30 queries to match the plan costs of QuickPick-1000.

Digging deeper, we found that the break-even point of 30 queries roughly corresponds to seeing all relations in the schema at least once. In fact, we can train \sys on small queries and test it on larger ones---as long as the relations are covered well.  To investigate this generalization power, we trained \sys on all queries with $\leq$ 9 and 8 relations, respectively, and tested on the remaining queries (out of a total of 113). For comparison we include a baseline scheme of training on 80 random queries and testing on 33; see Table~\ref{table:generalization}. 


Table~\ref{table:generalization} shows that even when trained on subplans, \sys performs relatively well and generalizes to larger joins (recall, the workload contains up to 15-way joins). This indicates that \sys indeed learns \emph{local structures}---efficient joining of small combinations of relations. When those local structures do not sufficiently cover the cases of interest during deployment, we see degraded performance.

\subsubsection{Relevance and Quality of Training Data}
\label{eval:data-relevance}
\begin{figure}[t!]
    \centering
    \includegraphics[width=0.9\columnwidth,keepaspectratio]{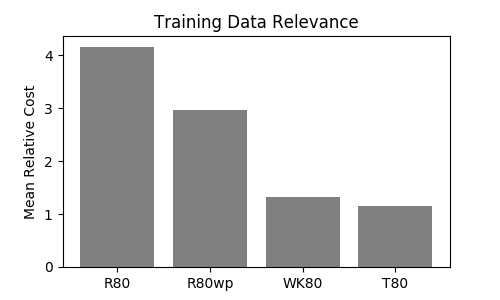}
\vspace{-0.4cm}
    \caption{\small{
    Relevance of training data vs. \sys's plan cost.
    {\sf R80} is a dataset sampled independently of the JOB queries with random joins/predicates from the schema.
    {\sf R80wp} has random joins as before but contains the workload's predicates. 
    {\sf WK80} includes 80 actual queries sampled from the workload. 
    {\sf T80} describes a scheme where each of the 33 query templates is covered at least once in sampling.  These schemes are increasingly ``relevant''. Costs are relative  w.r.t. \textbf{EX}.
    }\label{exp:plot3}}
\vspace{-0.2cm}
\end{figure}

Quantity of training data matters, and so do \textit{relevance} and \textit{quality}.  We first study relevance, i.e., the degree of similarity between the sampled training data and the test queries.  This is controlled by changing the training data sampling scheme. 
Figure \ref{exp:plot3} plots the performance of different data sampling techniques each with 80 training queries. It confirms that the more relevant the training queries can be made towards the test workload, the less data is required for good performance.

\begin{figure}[t!]
    \centering
    \includegraphics[width=\columnwidth]{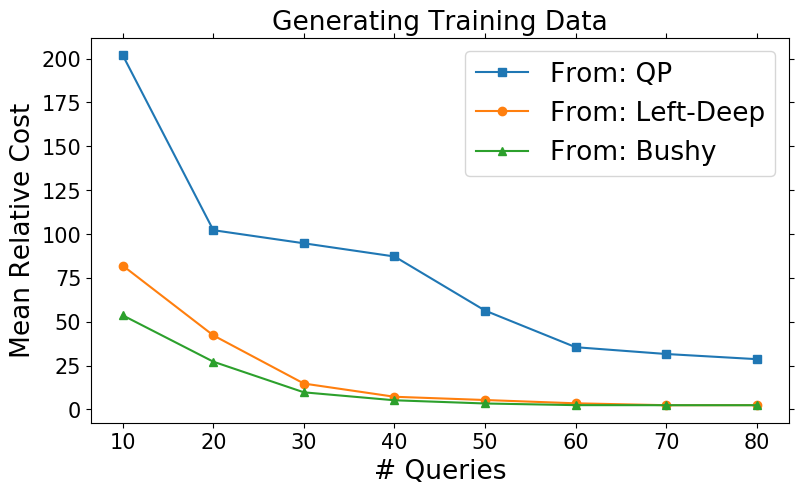}
\vspace{-0.7cm}
    \caption{
    \small{Quality of training data vs. \sys's plan cost.
    \sys trained on data collected from QuickPick-1000, left-deep, or the bushy (exhaustive) optimizer. Data variety boosts convergence speed and final quality. Costs are relative  w.r.t. \textbf{EX}.
    }\label{exp:plot4}}
\vspace{-0.4cm}
\end{figure}

\begin{table}[t!]\centering \small%
\begin{tabular}{@{} l c c @{}} \toprule
 & {\bf \# Training Queries}   & {\bf Mean Relative Cost}   \\ \midrule
{\bf Random} & 80 & 1.32   \\ 
{\bf Train} $\leq$ 9-way & 82 & 1.61 \\ 
{\bf Train} $\leq$ 8-way & 72 &  9.95  \\ 
 \bottomrule
\end{tabular}
\vspace{0.25em}
\caption{\small{\sys trained on small joins and tested on larger joins. Costs are relative to optimal plans.  }\label{table:generalization}}
\vspace{-0.7cm}
\end{table}

Notably, it also shows that even synthetically generated random queries ({\sf R80}) are useful. 
\sys still achieves a lower relative cost compared to QuickPick-1000 even with random queries (4.16 vs. 23.87).
This experiment illustrates that \sys does not actually require \emph{a priori} knowledge of the workload. 

Next, we study the \textit{quality} of training data, i.e., the optimality of the native planner \sys observes and gathers data from.  
We collect a varying amount of data sampled from the native optimizer, which we choose to be QuickPick-1000, left-deep, or bushy ({\bf EX}). 
Figure~\ref{exp:plot4} shows that all methods allow \sys to quickly converge to good solutions. The DP-based methods, left-deep and bushy, converge faster as they produce final plans and optimal subplans per query.  In contrast,  QuickPick  yields only 1000 random full plans per query.
The optimal subplans from the dynamic programs offer data variety valuable for training, and they cover better the space of different relation combinations that might be seen in testing.

\subsection{Real Systems Execution}
\label{eval:real-systems}
It is natural to ask: how difficult and effective is it for a production-grade system to incorporate \sys?  We address this question by integrating \sys into two systems, PostgreSQL and SparkSQL.\footnote{Versions: Spark 2.3; Postgres master branch checked out on 9/17/18.} The integrations were found to be straightforward: Postgres and SparkSQL each took less than 300 LoC of changes; in total about two person-weeks were spent. 

\subsubsection{Postgres Integration}
\label{sec:postgres-integration}
\sys integrates seamlessly with the bottom-up join ordering optimizer in Postgres. 
The original optimizer's DP table lookup is replaced with the invocation of 
\sys' Tensorflow (TF) neural network through the TF C API.  As discussed in \secref{eval:plan-latency}, plans are batch-evaluated to amortize the TF invocation overhead.  We run the Join Order Benchmark experiments on the integrated artifact and present the results below.  All of the learning utilizes the cost model and cardinality estimates provided by Postgres.

\vspace{0.5em} \noindent  \textbf{Training.} \sys observes the native cost model and cardinality estimates from Postgres. 
We configured Postgres
to consider bushy join plans (the default is to only consider left-deep plans). These plans generate traces of joins and their estimated costs in the form described in~\secref{sec:data-collection}. We \emph{do not} apply any exploration and execute the native optimizer as is. Training data is collected via Postgres' logging interface.  

Table~\ref{table:postgres-collection-overhead} shows that \sys can collect training data from an existing system with relatively minimal impact on its normal execution.  The overhead can be further minimized if training data is asynchronously, rather than synchronously, logged.

\begin{figure}[t]
    \centering
    \includegraphics[width=\columnwidth]{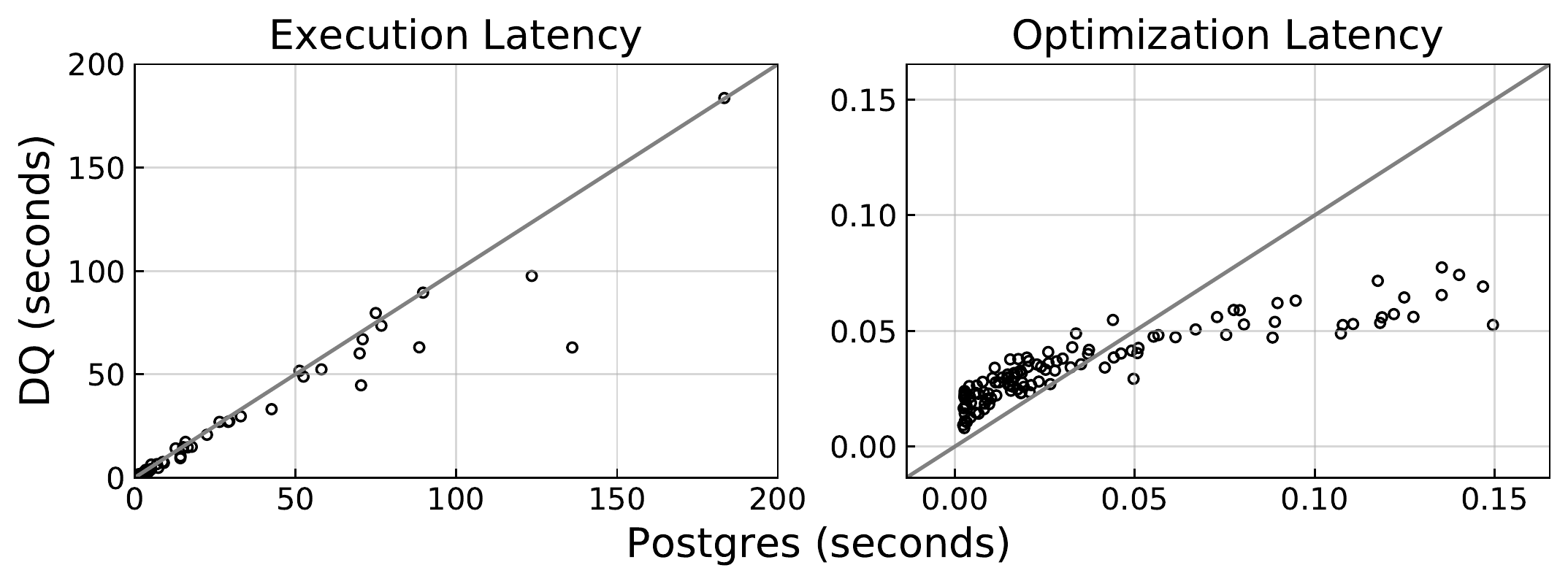}
\vspace{-0.8cm}
    \caption{\small {Execution and optimization latencies of \sys and Postgres on JOB. 
    Each point is a query executed by native Postgres (x-axis) and \sys (y-axis).
    Results below the $y=x$ line represent a speedup.  Optimization latency is the time taken for the full planning pipeline, not just join ordering.} \label{exp:real}} 
\vspace{-0.2cm}
\end{figure}

\vspace{0.5em} \noindent  \textbf{Runtimes on JOB (Figure~\ref{exp:real}).} We allow the Postgres query planner to plan over 80 of the 113 training queries. We use a 5-fold cross validation scheme to hold out different sets of 33 queries. Therefore, each query has at least one validation set in which it was unseen during training. We report the worst case planning time and execution time for queries that have multiple such runs.
In terms of optimization latency, \sys is significantly faster than Postgres for large joins, up to $3\times$. For small joins there is a substantial overhead due to neural network evaluations (even though \sys needs score much fewer join orders).  These results are consistent with the standalone experiment in Section~\ref{eval:plan-latency} and the same comments there on small-join regimes apply.
In terms of execution runtimes, \sys is significantly faster on a number of queries; averaging over the entire workload \sys yields a 14\% speedup.

\subsubsection{SparkSQL Integration}

\begin{table}[t]\centering \small%
\begin{tabular}{@{} l c c  @{}} \toprule
& {Median}  & {Max}  \\ \midrule
{\bf Postgres, no collection} & 19.17 ms  & 149.53 ms          \\ 
{\bf Postgres, with collection}  & 35.98 ms & 184.22 ms    \\ 
 \bottomrule
\end{tabular}
\vspace{0.25em}
\caption{\small{Planning latency with collection turned off/on.}}
\label{table:postgres-collection-overhead}
\vspace{-0.7cm}
\end{table}



\sys is also integrated into SparkSQL, a distributed data analytics engine.  To show that \sys's effectiveness applies to more than one workload, we evaluate the integrated result on TPC-DS.  

\vspace{0.5em} \noindent  \textbf{Training.} SparkSQL 2.3 contains a cost-based optimizer which enumerates bushy plans for queries whose number of relations falls under a tunable threshold.  We set this threshold high enough so that all queries are handled by this bushy dynamic program.  To score plans, the optimizer invokes \sys's trained neural net through TensorFlow Java.  We use the native SparkSQL cost model and cardinality estimates.  All algorithmic aspects of training data collection remain the same as the Postgres integration.

\begin{figure}[t]
    \centering
    \includegraphics[width=\columnwidth]{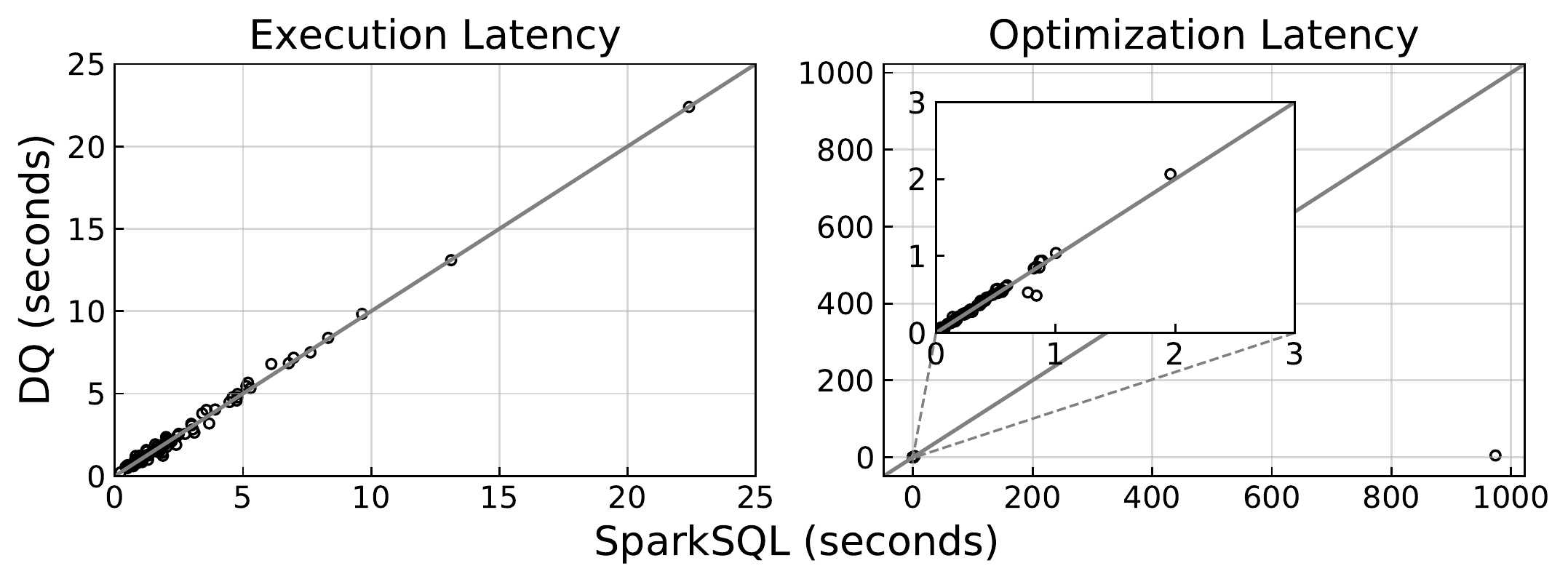}
    \vspace{-.8cm}
    \caption{\small{
      Execution and optimization latencies of \sys and SparkSQL on TPC-DS (SF1).  We use an EC2 c5.9xlarge instance with 36 vCPUs.  SparkSQL's bushy dynamic program takes 1000 seconds to plan the largest query (Q64, 18-relation join); we include a zoomed-in view of the rest of the planning latencies.  Results below the $y=x$ line represent a speedup.  Across the workload, \sys's mean speedup over SparkSQL for execution is 1.0$\times$ and that for optimization is 3.6$\times$.
    }} \label{fig:spark-tpcds}
\vspace{-0.3cm}
\end{figure}

\vspace{0.5em} \noindent  \textbf{Effectiveness on TPC-DS (Figure~\ref{fig:spark-tpcds}).}  We collect data from and evaluate on 97 out of all 104 queries in TPC-DS v2.4.  The data files are generated with a scale factor of 1 and stored as columnar Parquet files.  
  In terms of execution runtimes, \sys matches SparkSQL over the 97 queries (a mean speedup of 1.0$\times$).  In terms of optimization runtimes, \sys has a mean speedup of 3.6$\times$ but a max speedup of 250$\times$ on the query with largest number of joins (Q64).  
Note that the mean optimization speedup here is less drastic than JOB because TPC-DS queries contain much less relations to join. 

\vspace{0.5em} \noindent  \textbf{Discussion.} In summary, results above show that \sys's effective not only on the one workload designed to stress-test joins, but also on a well-established decision support workload.  Further, we demonstrate the ease of integration into production-grade systems including a RDBMS and a distributed analytics engine.  We hope these results provide motivation for developers of similar systems to incorporate \sys's learning-based join optimization technique.



\subsection{Fine-Tuning With Feedback}
\label{sec:eval-feedback}

Finally, we illustrate how \sys can overcome an inaccurate cost model by fine-tuning with feedback data (\secref{subsec:feedback}).  We focus on a specific JOB query, Q10c, where the cost model particularly deviates from the true runtime.  
Baseline \sys is trained on data collected over 112 queries, which is every query except for Q10c, as usual (i.e., values are costs from Postgres' native cost model).  For fine-tuning
we execute a varying amount of these queries and collect their actual runtimes.  To encourage observing a variety of physical operators, we use an exploration parameter of $\epsilon=0.1$ when observing runtimes (recall from \secref{sec:data-collection} exploration means with probability $\epsilon$ we form a random intermediate join).

\begin{figure}[tp]
    \centering
    \includegraphics[width=\columnwidth]{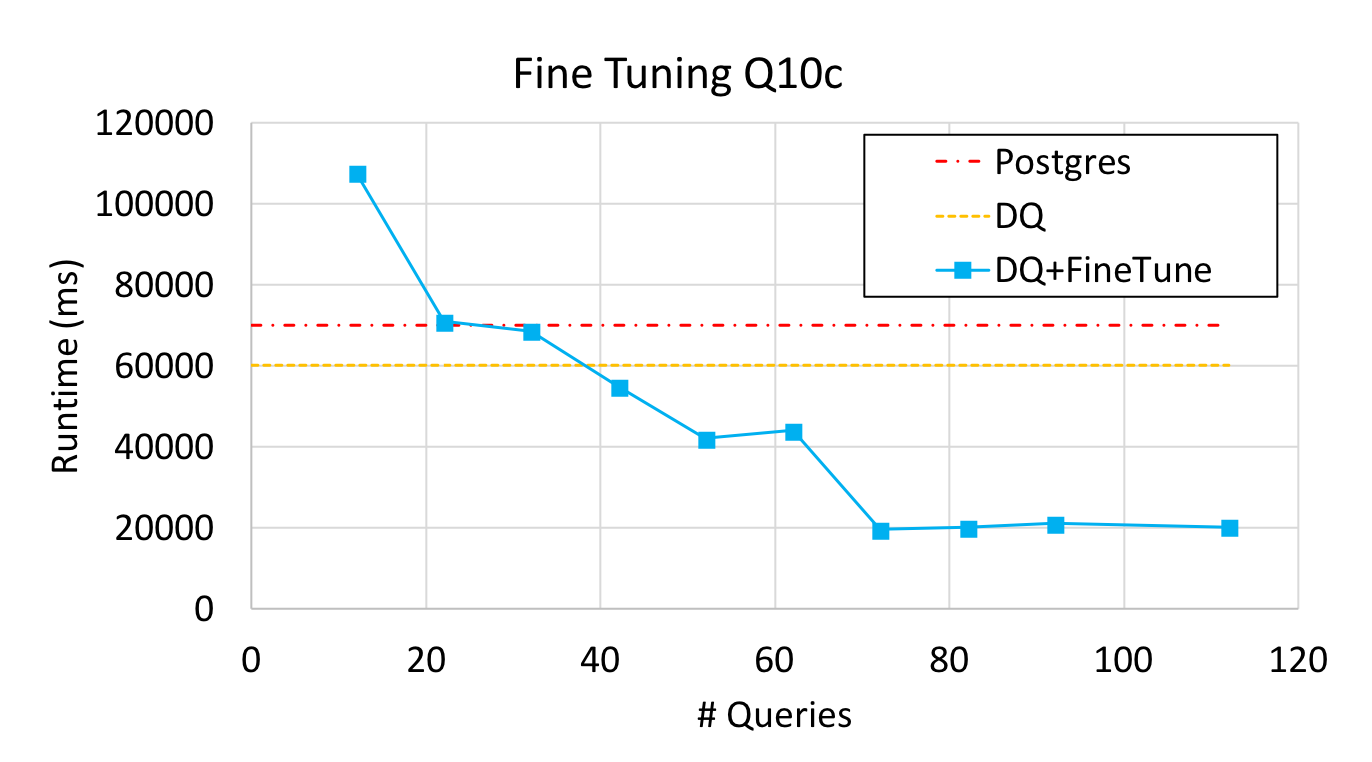}
    \vspace{-.7cm}
    \caption{Effects of fine-tuning \sys on JOB Q10c.  A modest amount of real execution using around 100 queries allows \sys to surpass both its original performance (by $3\times$) as well as Postgres (by $3.5\times$). \label{exp:fine-tuning}} 
\vspace{-0.3cm}
\end{figure}

Figure \ref{exp:fine-tuning} shows the results as a function of the number of queries observed for real execution.  Postgres emits a plan that executes in $70.0$s, while baseline \sys emits a plan that executes in $60.1$s.  After fine-tuning, \sys emits a plan that executes in $20.3$s, outperforming both Postgres and its original performance.  This shows true runtimes are useful in correcting faulty cost model and/or cardinality estimates.
Interestingly, training a version of \sys using only real runtimes failed to converge to a reasonable model---this suggests learning high-level features from inexpensive samples from the cost model is beneficial.


\section{Related Work}
\label{sec:relatedwork}
Application of machine learning in database internals is still the subject of significant debate this year and will continue to be a contentious question for years to come~\citep{btree, kraska2018case, mitzenmacher2018model, ma2018query}. An important question is what problems are amenable to machine learning solutions. We believe that query optimization is one such sub-area. The problems considered are generally hard and orders-of-magnitude of performance are at stake. In this setting, poor learning solutions will lead to slow but not incorrect execution, so correctness is not a concern.

\vspace{0.5em}\noindent \textbf{Cost Function Learning} 
We are certainly not the first to consider ``learning'' in the query optimizer and there are a number of alternative architectures that one may consider. The precursors to this work are attempts to correct query optimizers through execution feedback.
One of the seminal works in this area is the LEO optimizer~\citep{markl2003leo}. This optimizer uses feedback from the execution of queries to correct inaccuracies in its cost model. The underlying cost model is based on histograms. The basic idea inspired several other important works such as~\citep{chaudhuri2008pay}. The sentiment in this research still holds true today; when Leis et al. extensively evaluated the efficacy of different query optimization strategies they noted that feedback and cost estimation errors are still challenges in query optimizers~\citep{leis2015good}. A natural first place to include machine learning would be what we call \emph{Cost Function Learning}, where statistical learning techniques are used to correct or replace existing cost models. This is very related to the problem of performance estimation of queries~\citep{akdere2012learning, wu2013predicting, wu2013towards}. 

We actually investigated this by training a neural network to predict the selectivity of a single relation predicate. Results were successful, albeit very expensive from a data perspective. To estimate selectivity on an attribute with 10k distinct values, the training set had to include 1000 queries. This architecture suffers from the problem of \emph{featurization of literals}; the results are heavily dependent on learning structure in literal values from the database that are not always straightforward to featurize. This can be especially challenging for strings or other non-numerical data types.  A recent workshop paper does show some promising results in using Deep RL to construct a good feature representation of subqueries but it still requires $>$ 10k queries to train~\citep{ortiz2018learning}. 

\vspace{0.5em}\noindent \textbf{Learning in Query Optimization} 
Recently, there has been several exciting proposals in putting learning inside a query optimizer.  Ortiz et al.~\citep{ortiz2018learning} applies deep RL to learn a representation of queries, which can then be used in downstream query optimization tasks.  Liu~\citep{liu2015cardinality} and Kipf~\citep{kipf2018learned} use DNNs to learn cardinality estimates.  Closer to our work is Marcus et al.'s proposal of a deep RL-based join optimizer, ReJOIN~\citep{marcus2018deep}, which offered a preliminary
view of the potential for deep RL in this context. The early results reported in ~\citep{marcus2018deep} top 
out at a 20\% improvement in plan execution time of Postgres (compared to our 3x), and as of that paper they had only evaluated on 10 out of the 113 JOB queries that we study here.
DQ qualitatively goes beyond that work by offering an extensible
featurization scheme supporting physical join selection. More fundamentally, DQ integrates
the dynamic programming of Q-learning into that of a standard query optimizer, which allows us to use off-policy learning.
Due to use of on-policy
policy gradient methods, \citep{marcus2018deep} requires about 8,000 training
queries to reach native Postgres’ cost on the 10 JOB queries. DQ exploits
optimal substructures of the problem and uses off-policy
Q-learning to increase data-efficiency by two orders of magnitude:
80 training queries to outperform Postgres’
real execution runtimes on the entire JOB benchmark.

\vspace{0.5em}\noindent \textbf{Adaptive Query Optimization}
Adaptive query processing~\citep{avnur2000eddies,deshpande2007adaptive} as well as the related techniques to re-optimize queries during execution~\citep{markl2004robust, babu2005proactive} is another line of work that we think is relevant to the discussion. Reinforcement learning studies sequential problems and adaptive query optimization is a sequential decision problem over tuples rather than subplans.
We focus our study on optimization in fixed databases and the adaptivity that \sys offers is at a workload level. Continuously updating a neural network can be challenging for very fine-grained adaptivity, e.g., processing different tuples in different ways. 

\vspace{0.5em}\noindent \textbf{Robustness} There are a couple of branches of work that study robustness to different parameters in query optimization. In particular, the field of ``parametric query optimization''~\citep{hulgeri2002parametric,trummer2014multi}, studies the optimization of piecewise linear cost models. Interestingly, \sys is it is agnostic to this structure. It learns a heuristic from data  identifying different regimes where different classes of plans work. We hope to continue experiments and attempt to interpret how \sys is partitioning the feature space into decisions. There is also a deep link between this work and least expected cost (LEC) query optimization~\citep{chu2002least}. Markov Decision Processes (the main abstraction in RL) are by definition stochastic and optimize the LEC objective.

\vspace{0.5em}\noindent \textbf{Join Optimization At Scale}
Scaling up join optimization has been an important problem for several decades, most recently~\citep{neumann2018adaptive}. 
At scale, several randomized approaches can be applied.
There is a long history of randomized algorithms (e.g., the QuickPick algorithm~\citep{waas2000join}) and genetic algorithms~\citep{bennett1991genetic, steinbrunn1997heuristic}.
These algorithms are pragmatic and it is often the case that commercial optimizers will leverage such a method after the number of tables grows beyond a certain point.
The challenge with these methods is that their efficacy is hard to judge.
We found that QuickPick often varied in performance on the same query quite dramatically.

Another heuristic approach is relaxation, or solving the problem exactly under simplified assumptions.
One straightforward approach is to simply consider greedy search avoiding Cartesian products~\citep{fegaras1998new}, which is also the premise of the IK-KBZ algorithms~\citep{ibaraki1984optimal,krishnamurthy1986optimization}.
Similar linearization arguments were also made in recent work~\citep{trummer2017solving,neumann2018adaptive}.
Existing heuristics do not handle all types of non-linearities well, and this is exactly the situation where learning can help.
Interestingly enough, our proposed technique has a $O(n^3)$
runtime, which is similar to the \emph{linearizedDP} algorithm described in ~\citep{neumann2018adaptive}.
We hope to explore the very large join regime in the future and an interesting direction is to compare \sys to recently proposed techniques like ~\citep{neumann2018adaptive}.

\section{Discussion, Limitations, and Conclusion}
\label{sec:extensions}
We presented our method with a featurization designed for inner joins over foreign key relations as these were the major join queries in our benchmarks. This is not a fundamental restriction and is designed to ease exposition. 
It is relatively straightforward to extend this model to join conditions composed of conjunctions of binary expressions.  Assume the maximum number of expressions in the conjunction is capped at $\mathcal{N}$.
As before, let $A$ be the set of all attributes in the database. Each expression has two attributes and an operator. As with featurizing the vertices we can 1-hot encode the attributes present. We additionally have to 1-hot encode the binary operators $\{=,\neq,<,>\}$. 
 For each of the expressions in the conjunctive predicate, we concatenate the binary feature vectors that have its operator and attributes.  Since the maximum number of expressions in the conjunction capped at $\mathcal{N}$, we can get a fixed sized feature vector for all predicates. 
 
More broadly, we believe \sys is a step towards a learning query optimizer.
As illustrated by the Cascades optimizer~\citep{graefe1995cascades} and follow-on work, cost-based dynamic programming---whether bottom up or top-down with memoization---needs not be restricted to select-project-join blocks. 
Most query optimizations can be recast into a space of algebraic transformations amenable to dynamic programming, including asymmetric operators like outer joins, cross-block optimizations including order optimizations and ``sideways information passing'', and even non-relational operators like PIVOT.
The connection between RL and Dynamic Programming presented in this paper can be easily leveraged in those scenarios as well.
Of course this blows up the search space, and large spaces are ideal for solutions like the one we proposed.

It is popular in recent AI research to try ``end-to-end'' learning, where problems that were traditionally factored into subproblems (e.g., self-driving cars involve separate models for localization, obstacle detection and lane-following) are learned in a single unified model. 
One can imagine a similar architectural ambition for an end-to-end learning query optimizer, which simply maps subplan features to measured runtimes. This would require a significant corpus of runtime data to learn from, and changes to the featurization and perhaps the deep network structure we used here. \sys is a pragmatic middle ground that exploits the structure of the join optimization problem. Further exploring the extremes of learning and query optimization in future work may shed more insights.

{
\small
\bibliographystyle{abbrv}
\bibliography{ref,recent} 
\scriptsize
}

\appendix
\section{Standalone Optimization Experiment Setup}
\label{appendix:cost-details}
\noindent We consider three different cost models on the same workload:

\vspace{0.25em} \noindent \textbf{CM1: } In the first cost model (inspired by~\citep{leis2015good}), we model a main-memory database that performs two types of joins: index joins and in-memory hash joins. Let $O$ describe the current operator, $O_l$ be the left child operator, and $O_r$ be the right child operator. The costs are defined with the following recursions:
\[
\textsf{c}_{ij}(O) = \textsf{c}(O_l) + \textsf{match}(O_l, O_r) \cdot |O_l| 
\]
\[
\textsf{c}_{hj}(O) = \textsf{c}(O_l) + \textsf{c}(O_r) + |O|
\]
where \textsf{c} denotes the cost estimation function, $|\cdot|$ is the cardinality function, and \textsf{match} denotes the expected cost of an index match, i.e., fraction of records that match the index lookup (always greater than 1) multiplied by a constant factor $\lambda$ (we chose $1.0$). 
We assume indexes on the primary keys.
In this cost model, if an eligible index exists it is generally desirable to use it, since $\textsf{match}(O_l, O_r) \cdot |O_l|$ rarely exceeds $\textsf{c}(O_r) + |O|$ for foreign key joins.
Even though the cost model is nominally ``non-linear'', primary tradeoff between the index join and hash join is due to index eligibility and not dependent on properties of the intermediate results. For the JOB workload, unless $\lambda$ is set to be very high, hash joins have rare occurrences compared to index joins.   

\vspace{0.25em} \noindent \textbf{CM2: } In the next cost model, we remove index eligibility from consideration and consider only hash joins and nested loop joins with a memory limit $M$. The model charges a cost when data requires additional partitioning, and further falls back to a nested loop join when the smallest table exceeds the squared memory:

\scriptsize
\[
\textsf{c}_{join} =
\begin{cases}
\textsf{c}(O_l) + \textsf{c}(O_r) + |O|  ~~ \text{ if } |O_r| + |O_l| \le M\\
\textsf{c}(O_l) + \textsf{c}(O_r) + 2(|O_r|+|O_l|) + |O| ~~ \text{ if } \min(|O_r|, |O_l|) \le M^2\\
\textsf{c}(O_l) + \textsf{c}(O_r) + (|O_r| + 
\left \lceil{\frac{|O_r|}{M}}\right \rceil |O_l|)
\end{cases}
\]
\normalsize
The non-linearities in this model are size-dependent, so controlling the size of intermediate relations is important in the optimization problem.
We set the memory limit $M$ to $10^5$ tuples in our experiments. This limit is low in real-world terms due to the small size of the benchmark data. However, we intend for the results to be illustrative of what happens in the optimization problems.

\vspace{0.25em} \noindent \textbf{CM3: } In the next cost model, we model a database that accounts for the reuse of already-built hash tables. We use the Gamma database convention where the left operator as the ``build'' operator and the right operator as the ``probe'' operator~\citep{gerber1986data}. 
If the previous join has already built a hash table on an attribute of interest, then the hash join does not incur another cost.
\[
\textsf{c}_{nobuild} = \textsf{c}(O_l) + \textsf{c}(O_r) - |O_r| + |O|
\]
We also allow for index joins as in CM1.
This model makes hash joins substantially cheaper in cases where re-use is possible. This model favors some subplans to be right-deep plans which maximize the reuse of the built hash tables. Therefore, optimal solutions have both left-deep and right-deep segments.

\vspace{0.5em} In our implementation of these cost models, we use true cardinalities on single-table predicates, and we leverage standard independence assumptions to construct more complicated cardinality estimates. (This is not a fundamental limitation of \sys.  Results in ~\secref{eval:real-systems} have shown that when Postgres and SparkSQL provide their native cost model and cardinality estimates, \sys is as effective.) The goal of this work is to evaluate the join ordering process independent of the strength or weakness of the underlying cardinality estimation. 

We consider the following baseline algorithms. These algorithms are not meant to be a comprehensive list of heuristics but rather representative of a class of solutions.

\begin{table*}[ht]
\begin{tabular}{@{} l l l l l l l l l l l l l l @{}} \toprule
{\bf Optimizer}  && \multicolumn{3}{c}{\bf Cost Model 1}  & & \multicolumn{3}{c}{\bf Cost Model 2} & & \multicolumn{3}{c}{\bf Cost Model 3}\\
  && {Min}  & {Mean}  & {Max} && {Min}  & {Mean}  & {Max} && {Min}  & {Mean}  & {Max}\\ \midrule
QuickPick   (QP)&&  1    & 23.87 & 405.04  && 7.43  & 51.84  & 416.18 && 1.43 & 16.74 & 211.13  \\
IK-KBZ          (KBZ)&&  1    & 3.45  & 36.78   && 5.21  & 29.61  & 106.34 && 2.21 & 14.61 & 96.14   \\
Right-deep      (RD)&&   4.7  & 53.25 & 683.35  && 1.93  & 8.21   & 89.15  && 1.83 & 5.25  & 69.15   \\
Left-deep    (LD) &&   1    & 1.08  & 2.14    && 1.75  & 7.31   & 65.45  && 1.35 & 4.21  & 35.91   \\
Zig-zag         (ZZ) &&   1    & 1.07  & 1.87    && 1     & 5.07   & 43.16  && 1    & 3.41  & 23.13   \\
Exhaustive  (EX) &&   1    & 1     & 1       && 1     & 1      & 1      && 1    & 1     & 1       \\
DQ              &&        1    & 1.32  & 3.11    && 1     & 1.68   & 11.64  && 1    & 1.91  & 13.14   \\
Minimum Selectivity           (MinSel) &&    2.43 & 59.86 & 1083.12 && 23.46 & 208.23 & 889.7  && 9.81 & 611.1 & 2049.13 \\
IK-KBZ+DP    (LDP)&&     1    & 1.09  & 2.72    && 2.1   & 10.03  & 105.32 && 2.01 & 3.99  & 32.19  \\
\hline
\end{tabular}
\caption{Extended results including omitted techniques for all three cost models. \label{table:full-results}}
\end{table*}

\begin{enumerate}[noitemsep]
    \item Exhaustive (\textbf{EX}): This is a dynamic program that exhaustively enumerates all join plans avoiding Cartesian products.
    \item left-deep (\textbf{LD}): This is a dynamic program that exhaustively enumerates all left-deep join plans.
    \item Right-Deep (\textbf{RD}): This is a dynamic program that exhaustively enumerates all right-deep join plans.
    \item Zig-Zag (\textbf{ZZ}): This is a dynamic program that exhaustively enumerates all zig-zag trees (every join has at least one base relation, either on the left or the right)~\citep{ziane1993parallel}.
    \item IK-KBZ (\textbf{KBZ}): This algorithm is a polynomial time algorithm that decomposes the query graph into chains and orders the chains based on a linear approximation of the cost model~\citep{krishnamurthy1986optimization}. 
    \item QuickPick-1000 (\textbf{QP}): This algorithm randomly selects 1000 join plans and returns the best of them. 1000 was selected to be roughly equivalent to the planning latency of \sys~\citep{waas2000join}.
    \item Minimum Selectivity (\textbf{MinSel}): This algorithm selects the join ordering based on the minimum selectivity heuristic~\citep{neumann2018adaptive}. While MinSel was fast, we found poor performance on the 3 cost models used in the paper.
    \item Linearized Dynamic Program (\textbf{LDP}): This approach applies a dynamic program in the inner-loop of IK-KBZ~\citep{neumann2018adaptive}. Not surprisingly, LDP’s results  were highly correlated with those of IK-KBZ and Left-Deep enumeration, so we chose to omit them from the main body of the paper.
\end{enumerate}

All of the algorithms consider join ordering without Cartesian products, so \textbf{EX} is an optimal baseline. We report results in terms of the suboptimality w.r.t. \textbf{EX}, namely $cost_{algo}/cost_{\textbf{EX}}$.
We present results on all 113 JOB queries.
We train on 80 queries and test on 33 queries. We do 4-fold cross validation to ensure that every test query is excluded from the training set at least once.
The performance of \sys is only evaluated on queries not seen in the training workload.

Our standalone experiments are integrated with Apache Calcite~\citep{calcite}.
Apache Calcite provides libraries for parsing SQL, representing relational algebraic expressions, and a Volcano-based query optimizer~\citep{graefe1991volcano,graefe1995cascades}.  Calcite does not handle physical execution or storage and uses JDBC connectors to a variety of database engines and file formats.
We implemented a package inside Calcite that allowed us to leverage its parsing and plan representation, but also augment it with more sophisticated cost models and optimization algorithms.
Standalone \sys is written in single-threaded Java.
The extended results including omitted techniques are described in Table~\ref{table:full-results}.

\section{$C_{out}$ Cost Model}
We additionally omitted experiments with a simplified cost model only searching for join orders and ignoring physical operator selection. We fed in true cardinalities to estimate the selectivity of each of the joins, which is a perfect version of the ``$C_{out}$'' model. We omitted these results as we did not see differences between the techniques and the goal of the study was to understand the performance of DQ over cost models that cause the heuristics to fail. In particular, we found that threshold non-linearities as in CM3 cause the most problems.

\begin{table}[ht!]
\centering
\begin{tabular}{@{} l c @{}} \toprule
  $C_{out}$  & {\bf Mean}\\ \midrule
QP  & 1.02               \\
IK-KBZ     & 1.34               \\
LD  & 1.02               \\
ZZ    & 1.02               \\
Ex & 1                  \\
DQ         & 1.03               \\
MinSel     & 1.11  \\
\hline
\end{tabular}
\end{table}

\section{Additional Standalone Experiments}
In the subsequent experiments, we try to characterize when \sys is expected to work and how efficiently.

\subsection{Sensitivity to Training Data}
Classically, join optimization algorithms have been deterministic. Except for \textbf{QP}, all of our baselines are deterministic as well.
Randomness in \sys (besides floating-point computations) stems from what training data is seen.
We run an experiment where we provide \sys with 5 different training datasets and evaluate on a set of 20 hold-out queries. We report the max range (worst factor over optimal minus best factor over optimal) in performance over all 20 queries in Table~\ref{table:plan-variance}. 
For comparison, we do the same with \textbf{QP} over 5 trials (with a different random seed each time).


\begin{table}[h]\centering \small%
\begin{tabular}{@{} l c c c @{}} \toprule
    & {\bf CM1}  & {\bf CM2}  & {\bf CM3} \\ \midrule
{\bf QP}  & 2.11$\times$  & 1.71$\times$ & 3.44$\times$  \\
\sys  & 1.59$\times$ & 1.13$\times$ &  2.01$\times$\\
 \bottomrule
\end{tabular}
\vspace{0.25em}
\caption{\small{Plan variance over trials. \label{table:plan-variance}}}
\vspace{-0.6cm}
\end{table}

We found that while the performance of \sys does vary due to training data, the variance is relatively low. Even if we were to account for this worst case, \sys would still be competitive in our macro-benchmarks. It is also substantially lower than that of \textbf{QP}, a true randomized algorithm.

\subsection{Sensitivity to Faulty Cardinalities}
In general, the cardinality/selectivity estimates computed by the underlying RDBMS do not have up-to-date accuracy.  All query optimizers, to varying degrees, are exposed to this issue since using faulty estimates during optimization may yield plans that are in fact suboptimal.  It is therefore worthwhile to investigate this sensitivity and try to answer, ``is the neural network more or less sensitive than classical dynamic programs and heuristics?''


In this microbenchmark,  the optimizers are fed \emph{perturbed} base relation cardinalities (explained below) during optimization; after the optimized plans are produced, they are scored by an \emph{oracle} cost model.  This means, in particular, \sys only sees noisy relation cardinalities during training and is tested on true cardinalities. The workload consists of 20 queries randomly chosen out of all JOB queries; the join sizes range from 6 to 11 relations. The final costs reported below are the average from 4-fold cross validation.


The perturbation of base relation cardinalities works as follows. We pick $N$ random relations, the true cardinality of each is multiplied by a factor drawn uniformly from $\{2, 4, 8, 16\}$.  As $N$ increases, the estimate noisiness increases (errors in the leaf operators get propagated upstream in a compounding fashion).  Table~\ref{table:sensitivity-1} reports the final costs with respect to estimate noisiness.



\begin{table}[h]\centering \small%
\begin{tabular}{@{} l c c c c @{}} \toprule
& {$N=0$}  & {$N=2$}  & {$N=4$}    & {$N=8$} \\ \midrule
{\bf KBZ} & 6.33  & 6.35  & 6.35  & 5.85         \\ 
{\bf LD}  & 5.51 & 5.53  & 5.53   &  5.60         \\ 
{\bf EX}  & 5.51  & 5.53   & 5.53    & 5.60         \\ 
\sys & 5.68  & 5.70   & 5.96    & 5.68\\ 
 \bottomrule
\end{tabular}
\vspace{0.25em}
\caption{\small{Costs ($\log_{10}$) when $N$ relations have perturbed cardinalities.}}
\label{table:sensitivity-1}
\vspace{-0.6cm}
\end{table}

Observe that, despite a slight degradation in the $N=4$ execution, \sys  is not any more sensitive than the \textbf{KBZ} heuristic.  It closely imitates exhaustive enumeration---an expected behavior since its training data comes from \textbf{EX}'s plans computed with the faulty estimates.


\subsection{Ablation Study}
Table~\ref{table:feat-ablation} reports an ablation study of the featurization described earlier (\secref{subsec:featurization}): 


\begin{table}[h]\centering \small%
\begin{tabular}{@{} l c c c @{}} \toprule
    & {\bf Graph Features}  & {\bf Sel. Scaling}  & {\bf Loss} \\ \midrule
{\bf No Predicates}  & No  & No   & 0.087 \\ 
 & Yes  & No   & 0.049 \\ 
  & Yes  & Yes   & 0.049 \\ 
{\bf Predicates}  & No  & No   &  0.071\\ 
& Yes  & No   &  0.051\\ 
& Yes  & Yes   &  0.020\\ 
 \bottomrule
\end{tabular}
\vspace{0.25em}
\caption{\small{Feature ablation. \label{table:feat-ablation}}}
\vspace{-0.6cm}
\end{table}

Without features derived from the query graph (Figure~\ref{fig:query-graph-feat}) and selectivity scaling (Figure~\ref{fig:feat-sel-scaling}) the training loss is 3.5$\times$ more. These results suggest that all of the different features contribute positively for performance.

\begin{figure}
    \centering
    \includegraphics[width=\columnwidth,keepaspectratio]{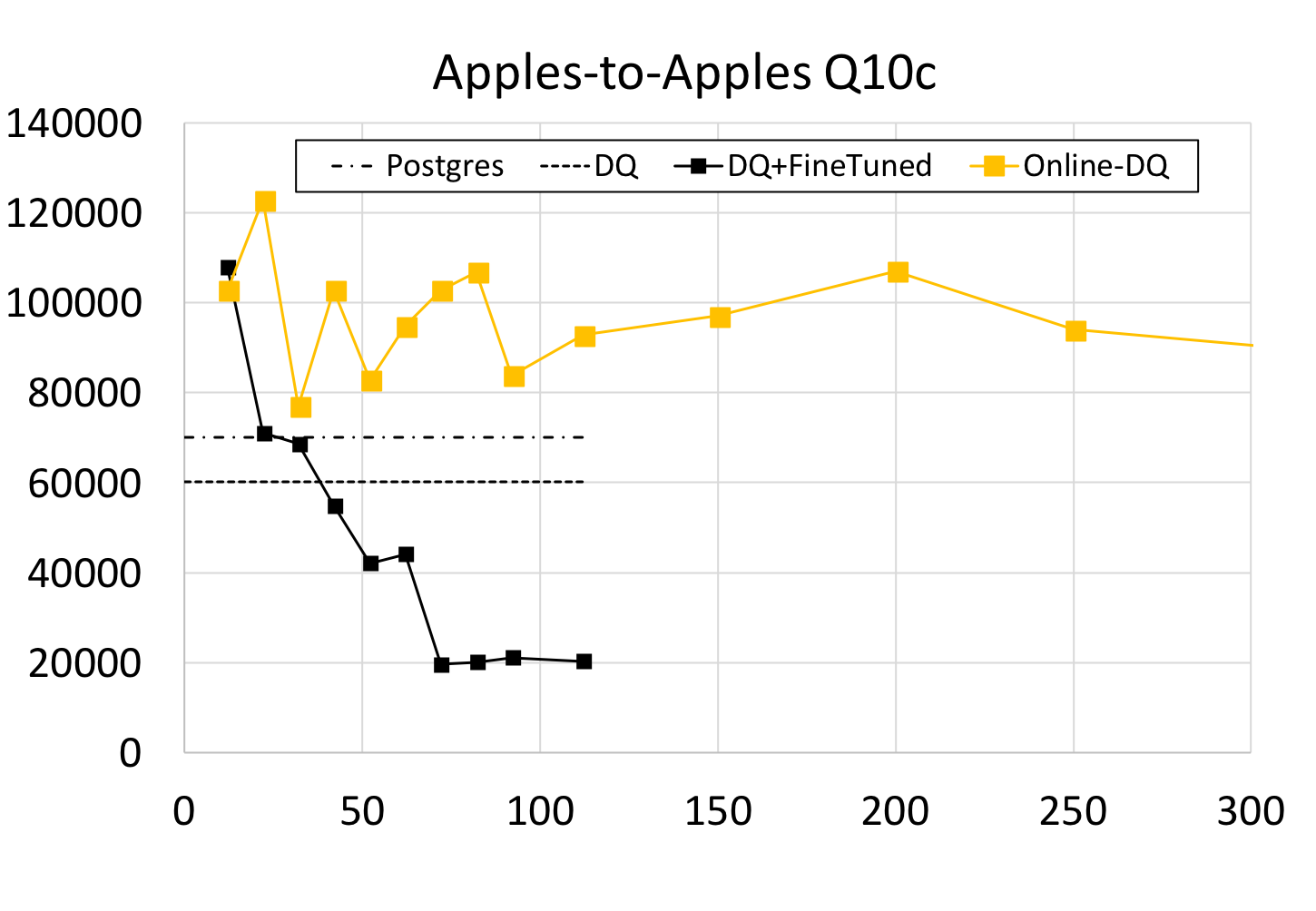}
    \caption{\small{We plot the runtime in milliseconds of a single query (q10c) with different variations of DQ (fully offline, fine tuning, and fully online). We found that the fine-tuned approach was the most effective one.} \label{exp:fine-tuning2}}
\end{figure}

\section{Discussion about Postgres Experiment}
We also run a version of DQ where the model is only trained with online data (effectively the setting considered in ReJOIN~\citep{marcus2018deep}). Even on an idealized workload of optimizing a single query (Query 10c), we could not get that approach to converge. We believe that the discrepancy from ~\citep{marcus2018deep} is due to physical operator selection. In that work, the Postgres optimizer selects the physical operators \textit{given} the appropriate logical plans selected by the RL policy. With physical operator selection, the learning problem becomes significantly harder (Figure~\ref{exp:fine-tuning2}).

We initially hypothesized the \sys outperforms the native Postgres optimizer in terms of execution times since it considers bushy plans. This hypothesis only partially explains the results.
We run the same experiment where \sys is restricted to producing left-deep plans; in other words, \sys considers the same plan space as the native Postgres optimizer. We found that there was still a statistically significant speedup:

\begin{table}[h]\centering \small%
\begin{tabular}{@{} l c c  @{}} \toprule
& {Mean}  & {Max}  \\ \midrule
 \sys {\bf:LD}  & 1.09$\times$ & 2.68$\times$    \\ 
\sys {\bf :EX}  & 1.14$\times$ & 2.72$\times$    \\ 
 \bottomrule
\end{tabular}
\vspace{0.25em}
\caption{\small{Execution time speedup over Postgres with different plan spaces considered by \sys. Mean is the average speedup over the entire workload and max is the best case single-query speedup.}}
\label{table:postgres-speedup}
\vspace{-0.6cm}
\end{table}

We speculate that the speedup is caused by imprecision in the Postgres cost model. As a learning technique, \sys may smooth out inconsistencies in the cost model. 

Finally, we compare with Postgres' genetic optimizer (GEQ) on the 10 largest joins in JOB. \sys is about 7\% slower in planning time, but nearly 10$\times$ faster in execution time. The difference in execution is mostly due to one outlier query on which GEQ is 37$\times$ slower.

\end{document}